\documentclass[a4paper,fleqn,usenatbib]{mnras}
\pdfoutput=1

\usepackage{graphicx}
\usepackage{epsfig}
\usepackage{rotating}
\usepackage{amssymb}
\usepackage{color}
\usepackage{fancyhdr}
\usepackage{textcomp}
\usepackage{longtable}
\usepackage{tabularx}
\usepackage{url}
\usepackage{graphics}

\begin{document}
\title[Non-linearity and environmental dependence of the star forming galaxies Main Sequence]{ Non-linearity and environmental dependence of the star forming galaxies Main Sequence}
\author[Erfanianfar dvi.]{G. Erfanianfar $^{1,2}$\thanks{E-mail:
erfanian@mpe.mpg.de}, P. Popesso$^{1,2}$, A. Finoguenov$^{3}$, D. Wilman$^{2}$, S. Wuyts$^{2,4}$,  A. Biviano$^{5}$ ,
\newauthor  M.~Salvato${^2}$, M. Mirkazemi$^1$, L.~Morselli${^1}$, F. Ziparo$^6$,  K.~Nandra$^{^2}$,  D.~Lutz$^{2}$, D.~Elbaz $^{7}$
\newauthor , M.~Dickinson$^{8}$, M. ~Tanaka$^{9}$, M. B.~Altieri$^{10}$, H. Aussel$^{7}$,   F. ~Bauer$^{11,12}$,
\newauthor  S. Berta$^2$, R. M. Bielby $^{13}$, N. Brandt$^{14}$, N. Cappelluti$^{15}$, A. Cimatti$^{16}$, M. C. Cooper$^{17}$,
\newauthor   D. Fadda$^{18}$,  O. Ilbert$^{19}$, E. Le Floch$^{7}$, B. Magnelli$^{20}$, J. S. Mulchaey$^{21}$, R. Nordon$^{22}$,
\newauthor   J. A. Newman$^{23}$,  A. Poglitsch${^2}$, F. Pozzi$^{16}$\\
$^{1}$Excellence Cluster Universe, Boltzmannstr. 2, 85748 Garching bei M\"{u}nchen, Germany\\
$^{2}$Max-Planck-Institut f\"{u}r extraterrestrische Physik, Giessenbachstra\ss e 1, 85748 Garching bei M\"{u}nchen, Germany\\
$^{3}$University of Helsinki, Department of Physics, P.O. Box 64, FI-00014 Helsinki \\
$^{4}$Department of Physics, University of Bath, Claverton Down, Bath, BA2 7AY, UK  \\
$^{5}$INAF/Osservatorio Astronomico di Trieste, via G. B. Tiepolo 11, 34131 Trieste, Italy\\
$^{6}$School of Physics and Astronomy, University of Birmingham, Edgbaston, Birmingham B15 2TT, UK\\
$^{7}$Laboratoire AIM, CEA/DSM-CNRS-Universit{\'e} Paris Diderot, IRFU/Service d'Astrophysique,  B\^at.709, CEA-Saclay,\\
 91191 Gif-sur-Yvette Cedex, France.\\
$^{8}$National Optical Astronomy Observatory, 950 North Cherry Avenue, Tucson, AZ 85719, USA\\
$^{9}$National Astronomical Observatory of Japan, 2-21-1 Osawa, Mitaka, Tokyo 181-8588, Japan\\
$^{10}$Herschel Science Centre, European Space Astronomy Centre, ESA, Villanueva de la Ca\~nada, 28691 Madrid, Spain\\
$^{11}$Instituto de Astrof́ısica, Facultad de F́ısica, Pontificia Universidad Catolica de Chile, 306, Santiago 22, Chile\\
$^{12}$Space Science Institute, 4750 Walnut Street, Suite 205, Boulder, CO 80301, USA\\
$^{13}$Dept. of Physics, Durham University, South Road, Durham, DH1 3LE, UK\\
$^{14}$Department of Astronomy and Astrophysics, 525 Davey Laboratory, The Pennsylvania State University, University Park, PA 16802, USA\\
$^{15}$INAF-Osservatorio Astronomico di Bologna, Via Ranzani 1, I-40127 Bologna, Italy\\
$^{16}$Dipartimento di Astronomia, Universit{\`a} di Bologna, Via Ranzani 1, 40127 Bologna, Italy\\
$^{17}$Center for Cosmology, Department of Physics and Astronomy, 4129 Reines Hall, University of California, Irvine, CA 92697, USA\\
$^{18}$NASA Herschel Science Center, Caltech 100-22, Pasadena, CA 91125, USA\\
$^{19}$LAM, CNRS-UNiv Aix-Marseille, 38 rue F. Joliot-Curis, F-13013 Marseille, France\\
$^{20}$Argelander-Institut f\"{u}r Astronomie, Universit\"{a}t Bonn, Auf dem H\"ugel 71, D-53121 Bonn, Germany\\
$^{21}$The Observatoires of the Carnegie Institution of Science, 813 Santa Barbara Street, Pasadena, CA 91101, USA\\
$^{22}$School of Physics and Astronomy, The Raymond and Beverly Sackler Faculty of Exact Sciences, Tel Aviv University, Tel Aviv 69978, Israel\\
$^{23}$Department of Physics and Astronomy, University of Pittsburgh and PITT-PACC, 3941 O’Hara St., Pittsburgh, PA 15260, USA\\
}
\maketitle

\date{Accepted 2015 October 23.  Received 2015 October 16; in original form 2015 June 25.}

\maketitle

\label{firstpage}
\clearpage
\begin{abstract}
Using data from four deep fields (COSMOS, AEGIS, ECDFS, and CDFN), we study the correlation between the position of galaxies in the star formation rate(SFR) versus stellar mass plane and local environment at $z<1.1$. To accurately estimate the galaxy SFR, we use the deepest available Spitzer/MIPS 24 and Herschel/PACS datasets. We distinguish group environments ($M_{halo}\sim$10$^{12.5-14.2}$$M_{\odot}$) based on the available deep X-ray data and lower halo mass environments based on the local galaxy density. We confirm that the Main Sequence (MS) of star forming galaxies is not a linear relation and there is a flattening towards higher stellar masses ($M_*>10^{10.4-10.6}$ $M_{\odot}$), across all environments. At high redshift  ($0.5<z<1.1$), the MS varies little with environment. At low redshift ($0.15<z<0.5$), group galaxies tend to deviate from the mean MS towards the region of quiescence with respect to isolated galaxies and less-dense environments. We find that the flattening of the MS toward low SFR is due to an increased fraction of bulge dominated galaxies at high masses. Instead, the deviation of group galaxies from the MS at low redshift is caused by a large fraction of red disk dominated galaxies which are not present in the lower density environments. Our results suggest that above a mass threshold ($\sim10^{10.4}-10^{10.6}$$M_{\odot}$) stellar mass, morphology and environment act together in driving the evolution of the SF activity towards lower level. The presence of a dominating bulge and the associated quenching processes are already in place beyond $z\sim$1. The environmental effects appear, instead, at lower redshifts and have a long time-scale.

\end{abstract}

 \section{Introduction}
 
The so-called main sequence (MS) of star forming galaxies is a tight correlation between the star formation rates (SFRs) and the stellar masses of the bulk of the star forming galaxy population. The MS appears to hold at least over the past 10 Gyr (\citealt{Noeske2007a, Elbaz2007, Daddi2007, Peng2010, Whitaker2012, Whitaker2014}).  The zero point of this relation evolves with cosmic time such that the level of star formation activity was much higher in the past. Indeed, there is an uncontroversial steep decline in the cosmic star formation history (CSFH) by a factor of 10 since z $\sim$ 1, after a peak of activity around z $\sim$ 1--2 (\citealt{Lilly1996, Madau1998,  LeFloch2005, Hopkins2006}).  Aside from the normalization, the slope and the dispersion of the relation are still quite uncertain. The measurement of the slope and dispersion vary widely in the literature, ranging between 0.2--1.2 for the slope and 0.3--0.6 for the dispersion (see \citealt{Speagle2014} and references therein). Most of the uncertainty is due either to biases introduced by the selection criteria used to identify the star forming galaxy population, SFR indicators, method of fitting (mean, median, mode)  or because  the relation is not linear but its slope varies at different mass scales. Indeed, \cite{Whitaker2014} show that at 0.5 $< z <$ 2, the MS exhibits a steeper slope in the low stellar mass regime (stellar masses lower than $10^{10}$ $M_{\odot}$) than at higher masses.

A detailed understanding of the evolution of the MS is essential to encode fundamental information of the galaxy growth. Indeed, while the slope indicates how the star formation activity varies at different stellar mass scales, the dispersion of this relation reveals the level of stochasticity in the gas accretion history and the star formation efficiency (\citealt{Leitner2012};\citealt{Behroozi2013}). Thus, understanding how galaxies move as a function of time in the SFR-stellar mass plane with respect to the MS is a very powerful tool to identify the processes that lead to the progressive suppression of the star formation activity as a function of time. 

In a schematic view, such processes can be divided into two big categories: ``internal" vs. ``external" processes.  Currently it is still heavily debated to what extent the properties of galaxies are determined by each of these two scenarios. In the current paradigm of galaxy formation the ``internal" processes are mainly linked to the co-evolution of the host galaxy and its central black hole (\citealt{ Croton2006, DeLucia2006}). Most of the galaxy formation models recognize Active Galactic Nuclei (AGN) feedback as the main mechanism to drive the gas away and stop the growth of the galaxy and its central black hole (BH).  Two main mechanisms are proposed: ``quasar-mode" active galactic nucleus (AGN) feedback  and ``radio-mode" AGN feedback (e.g., \citealt{Croton2006}). While these two mechanisms could be effective for very massive galaxies ( at stellar masses above $10^{11}$ $M_{\odot}$, \citealt{Genzel2014}), such feedbacks are not observed for the bulk of the star forming galaxy population (\citealt{Rovilos2012, Mullaney2012, Bongiorno2012, Rosario2012, Harrison2012}). Alternatively, \citealt{Martig2009} propose that ``morphological quenching'' may ``internally'' switch off or reduce the efficiency of star formation in galaxies through the formation of a dominant bulge that stabilizes the gas disk against gravitational instabilities (see \citealt{Saintonge2012} and \citealt{Crocker2012} for observational hints in the local universe, and \cite{Genzel2014} for galaxies at high redshift). 

The ``external'' processes are, instead, linked to either the interaction with other galaxies or with the local environment. While mergers  (\citealt{Park2009}), tidal gas stripping  (\citealt{Dekel2003, Diemand2007}) and harassment (\citealt{Farouki1981, Moore1998}) are included in the former class of external processes, a plethora of different mechanisms are counted in the latter group, such as strangulation/starvation (\citealt{Larson1980}), ram pressure stripping (\citealt{Gunn1972, Abadi1999}), turbulence and viscosity (e.g. \citealt{Quilis2001}). In addition, according to the accretion theory(e.g., \citealt{Birnboim2003, Keres2005}),  dark matter haloes exceeding a critical mass of  $M_{halo} \sim$10$^{12}$ $M_{\odot}$ should be able to stop the flow of infalling cold gas onto their central galaxies via virial shock heating. This mechanism, which is called more generally ``halo quenching'', would lead to a decrease or suppression of star formation over longer timescales. However, this process should be effective $z <1$, because at higher redshift the gas is predicted to penetrate to the central galaxy through cold streams even in massive halos (e.g., \citealt{Dekel2009}).

Both types of processes, internal and external, have been largely studied in the literature to identify the mechanism or the combination of them responsible for the suppression of the SF activity in galaxies and, thus, for their evolution on the SFR-stellar mass plane. \cite{Noeske2007a} and \cite{Wuyts2011} find evidence for a remarkable correlation between galaxy structure or morphology and the location of galaxies in the SFR-stellar mass plane. The MS would coincide with a morphological sequence of late type galaxies. This correlation is found to be in place already three billion years after the Big Bang. In addition, recently, \cite{Lang2014}, using deep high-resolution imaging of all five CANDELS fields, show that there is an increase in typical S\'ersic index and bulge to total ratio among star forming galaxies above 10$^{11}$ $M_{\odot}$ over the 0.5 $<z<$ 2.5 redshift range. They suggest that star formation quenching process must be internal and strongly linked with the bulge growth. However, on the other side, the existence of the so called morphology-density relation could set in turn a link between the location of a galaxy in the SFR-stellar mass plane and the environment where the galaxy lives. Indeed, since early type galaxies mainly populate high density regions (groups and clusters) and late type galaxies are generally more isolated, the relation between the morphology and the galaxy location with respect to the MS may translate into a relation with the environment. Thus, to build a global picture of galaxy evolution, it is mandatory to understand how the evolution of galaxies and their environment is intertwined. In this respect, there is still a lack of comprehensive study of the position of galaxies relative to the MS in different environments. 

In this paper, we intend, first, to examine whether the position the MS depends on the environment without applying pre-selection in defining the star forming galaxy population. Second we analyse the evolution of the galaxies in different environments with respect to the MS to understand if ``external processes'' can play a role in the evolution of the galaxy SF activity.  We define the MS simply by looking at the distribution of galaxies in the SFR-stellar mass plane in several stellar mass bins and as a function of the environment. This allows us to analyse the linearity of the MS as a function of the stellar mass and to study the MS dispersion in relation to the environment. For this purpose we use the largest available X-ray selected galaxy group sample created in \cite{Erfanianfar2014}, E14 hereafter. This sample, which comprises 88 galaxy groups at $0.1 <z < 1.1$ and down to a total mass of  $10^{12.5}$ $M_{\odot}$, is defined on the basis of the X-ray observations with Chandra and XMM of the major blank fields, such as COSMOS, AEGIS, ECDFS and CDFN. Isolated galaxies are instead defined on the basis of their local galaxy overdensity. These fields combine deep photometric (from the X-rays to the far-infrared wavelengths) and spectroscopic (down to i$_{AB}$ $\sim$ 24 mag and b $\sim$ 25 mag) observations over relatively large areas to lead, for the first time, to the construction of statistically significant samples of groups up to high redshift with a secure spectroscopic galaxy membership (see $\S$2.2.2 and $\S$3.1 in E14). In addition, we use the latest and deepest available Spitzer MIPS and Herschel PACS (Photoconduct- ing Array Camera and Spectrometer, \citealt{Poglitsch2010}) mid and far infrared surveys, respectively, conducted on the same blank fields to retrieve an accurate measure of the star-formation rate of individual group galaxies.

The paper is organized as follows. In Sect. 2 we describe our dataset and how all relevant quantities are estimated.
 In Sect. 3 we describe our results and in Sect. 4 we discuss them and draw our conclusions. We adopt a \cite{Chabrier2003}
 initial mass function (IMF), H$_0=71$ km~s$^{-1}$~Mpc$^{-1}$, $\Omega_m=0.3$ and $\Omega_{\Lambda}=0.7$ throughout this paper.

 \section{Data}

The aim of this work is to study the evolution of the location of galaxies in the SFR-stellar mass plane as a function of the environment. Three main ingredients are necessary for such analysis: an accurate estimate of the galaxy SFR and stellar mass and a proper definition of the environment. For this purpose, we use the best studied blank fields as the AEGIS field, COSMOS, the ECDFS and the CDFN, to build a multi-wavelength dataset which combines deep photometry from the UV to far infrared wavelength to estimate accurately the galaxy SFR and stellar mass, and deep X-rays observations and high spectroscopic coverage to study the galaxy environment. A full description of the data available in each field is provided in $\S$2.1 of  E14. In this work we summarize how the galaxy SFR and stellar mass is estimated and how the environment is defined.

\subsection{Star formation rate and stellar masses}
 
To accurately estimate the galaxy star formation rate, we use the deepest available Spitzer MIPS 24 $\mu$m and PACS 100 and 160 $\mu$m datasets for all considered fields (see $\S$2.3 of E14 for details). We fit the available infrared data with the SED templates from \cite{Elbaz2011}. The total galaxy IR luminosity is estimated by integrating the best fit SED from 8-1000 $\mu$m. The SFR is, then, derived from the IR luminosity by using the \cite{Kennicutt1998} relation converted for a Chabrier IMF. 
However, due to the flux limits of the MIPS and PACS catalogs, the IR catalogs are sampling only the Main Sequence region and can not provide a SFR estimate for galaxies below the Main Sequence or in the region of quiescence. For a complete census of the star formation activity, we need an estimate of the SFR for all galaxies. For this reason, we complement the SFR estimates derived from IR data (SFR$_{IR}$), with a SFR based on the SED fitting technique (SFR$_{SED}$, \citealt{Wuyts2011} for AEGIS field and for CDFN, \citealt{Ilbert2010} for the COSMOS field and \citealt{Ziparo2013} for ECDFS). After converting the SFR$_{SED}$  and the stellar masses of each catalog to the same Chabrier IMF, we calibrate the SFR$_{SED}$  using the available SFR$_{IR}$ and stellar masses in each field (see $\S$2.3.1 of E14 for details). We apply this calibration only to galaxies with SFR$_{SED}$$>$-0.5, which is the limits of the available mid and far-infrared derived SFR. This calibration leads to a SFR$_{SED}$  estimate that is consistent with the SFR$_{IR}$ with a scatter of 0.3 dex over the stellar mass range $10^9-10^{12}$ $M_{\odot}$. The reader is referred to $\S$2.3.1 of E14 for the details about the redshift dependence of such calibration and the analysis of the scatter of the SFR$_{SED}$-SFR$_{IR}$ relation as a function of stellar mass. As reported in E14, the stellar masses estimated in the different catalogs are instead consistent with a scatter of $\sim$ 0.2 dex after the correction to the same IMF. We, thus, use the stellar masses provided by the mentioned catalogs, converted to the Chabrier IMF. We note that \cite{Ziparo2013}, using similar datasets, have already established the consistency of the SFRs and stellar masses derived from different methods (see $\S$3.3 of \citealt{Ziparo2013} for details).

\subsection{The environment}
\label{env}

Most of the literature on the role of environment in galaxy evolution defines the environment through the local galaxy density field (e.g. \citealt{Peng2010, Peng2012}).  According to simulations, the galaxy density distribution should trace the mass density field with a {\it{bias}} that depends on the galaxy stellar mass, since massive systems tend to be more clustered while low mass objects are more uniformly distributed. Nevertheless, the accuracy with which the local galaxy density field can be reconstructed has to cope with observational limitations. We adopt  the following approach to define the environment: we use the X-ray observations available in each field to identify massive halos through the X-ray emission of their intra-group or intra-cluster medium down to the group regime. The spectroscopic information is, then, used to define the group and cluster membership of all secure X-ray detected structures through dynamical analysis. With this method we create our group galaxy sample. By using the same spectroscopic redshift information, we define also the galaxy density field and we use it to identify isolated galaxies, that, according to the simulation are embedded in low mass halos. In addition we use the density indicator to identify galaxies at the same density of the group galaxies but not associated to any secure X-ray emission. Those galaxies should lie, thus, in filaments or in lower mass halos, where the local galaxy density can be the same as in massive groups due to projection effects. We will define this sample as ``filament-like'' galaxies. The comparison of the location of filament like and group galaxies, in particular, in the SFR-stellar mass plane, will reveal whether relative vicinity of galaxies (expressed through the density) or rather the halos in which galaxies are embedded (expressed through the group total mass) is most strongly impacting the evolution of the galaxy star formation activity.

In the following subsections we provide a more detailed description of the group, filament-like and isolated galaxy samples, respectively.
 
\subsubsection{Group galaxies}
 
For the purpose of this work, we use the group galaxy sample defined in E14. Briefly, we use a sample of member galaxies of X-ray selected groups drawn from the ECDFS, CDFN, AEGIS and COSMOS X-ray surveys (see \citealt{Finoguenov2015} for the detailed discussion of the X-ray detection limits). The group sample comprises 83 groups with halo masses ($M_{200}$) ranging from $10^{12.5}$ to $10^{14.3}$ $M_{\odot}$ in the redshift range [0.15$-$1.1]. All groups are spectroscopically confirmed and clearly identified as X-ray extension and galaxy overdensity along the line of sight (see $\S$ 2.2 of E14 for more details about the optical identification of the X-ray extended sources). The group membership is obtained via dynamical analysis. As explained in $\S$2.2.2 of E14, first we determine the group velocity dispersion ($\sigma$). This is used, then, to define the group membership within 2 $\times$ r$_{200}$ (the radius enclosing the group mass $M_{200}$) and with recessional velocity within 3 times the estimate of the group velocity dispersion. In order to follow the evolution of the position of group galaxies with respect to the MS, we divide the sample of group galaxy members in two redshift bins: one at low redshift (0.15 $<z<$ 0.5) and one at high redshift (0.5 $<z<$ 1.1) bin. In total we have 424 member galaxies in the low redshift bin and 511 member galaxies in the high redshift bin.

 \subsubsection{Field and filament-like galaxies : the local galaxy density}

Galaxies in dark matter halos with mass below $10^{12}$ $M_{\odot}$ can be clearly identified at the lowest values of the density field estimator. We use the galaxy density estimator which is provided by the galaxy number density of systems with a stellar mass above $10^{10}$ $M_{\odot}$, measured within a cylinder with radius of 0.75 Mpc and a velocity range of 1000 $\rm{km/s}$ around each galaxy. This type of density estimator takes advantage of the mass segregation observed in more massive halos (\citealt{Scodeggio2009}) and it is defined to sample a volume (0.75 Mpc and $\pm$1000 $\rm{km/s}$) quite similar to the one of a group/cluster sized halo. Indeed, the $r_{200}$ radius varies from 0.3-0.5 Mpc for a group to 1-1.5 Mpc for a massive cluster and 1000 $\rm{km/s}$ is roughly twice the velocity dispersion of a group and quite similar to the velocity dispersion of a massive cluster. The galaxy density derived with this approach is further corrected for the spectroscopic incompleteness that would lead to an underestimation of the actual galaxy density.  This correction is estimated with the same approach described in $\S$2.4 of E14. Briefly, we estimate the ratio between the number of all galaxies and those with known spectroscopic redshift with mass above $10^{10}$ $M_{\odot}$, in a cylinder along the line of sight of the considered galaxy with radius corresponding to 0.75~Mpc at the redshift of the considered source and with $|z_{source}-z_{phot}|<10000$ $\rm{km/s}$, where $z_{source}$ is the redshift of the central source and $z_{phot}$ is the photometric redshift of the surrounding galaxies.  The limit of $10000$ $\rm{km/s}$ is roughly 3 times the error of the photometric redshifts (\citealt{Ilbert2010}).  Of course $z_{phot}$ is replaced by the spectroscopic redshift whenever this is available.

\begin{figure}
  \centering
  \includegraphics[width=0.4\textwidth]{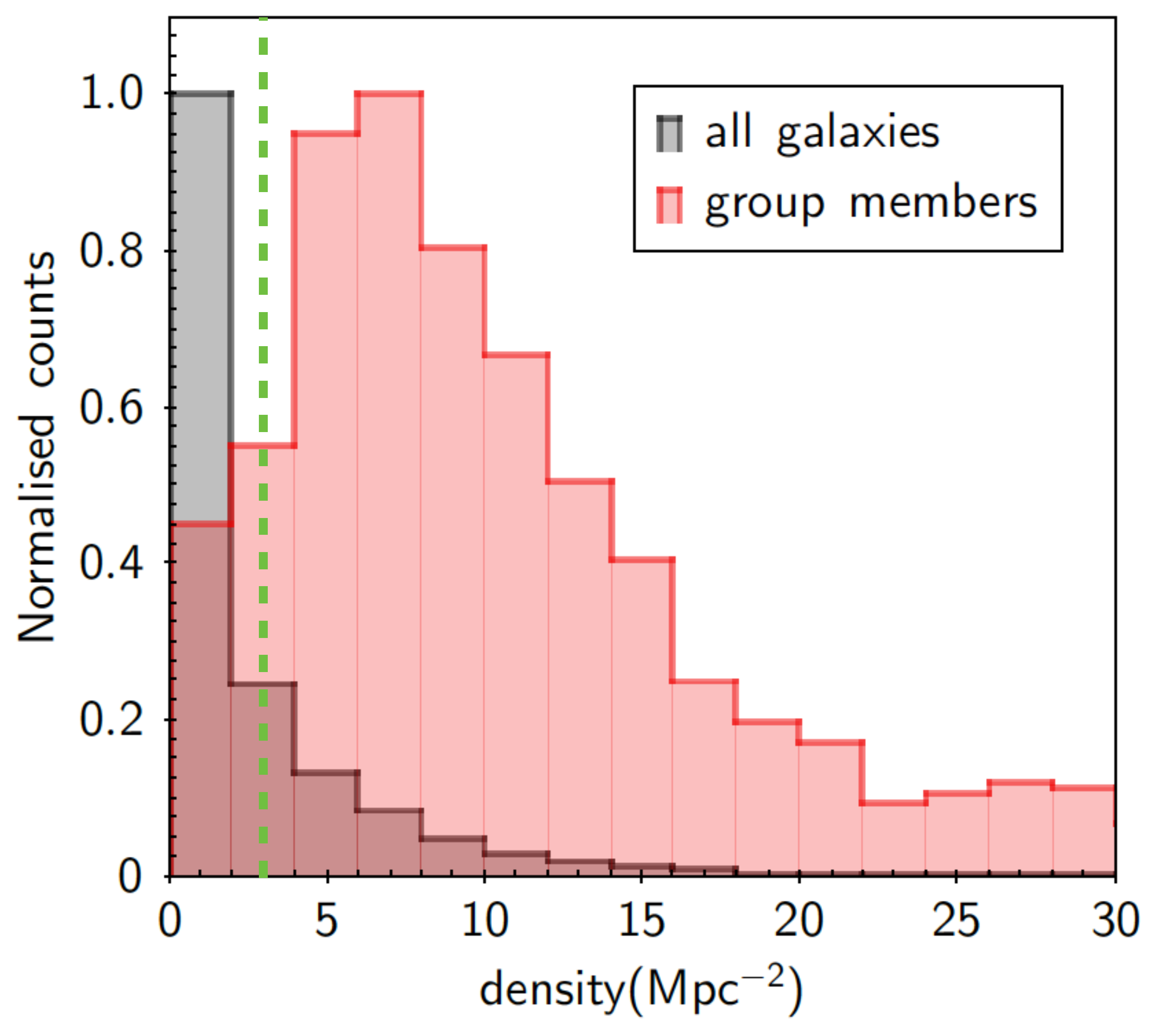}
  \caption[Density distribution around each galaxy with spectroscopic redshift]{Density distribution around each galaxy with spectroscopic redshift in AEGIS, COSMOS, ECDFS and CDFN (in black) and group member galaxies
in four mentioned fields (in red). The green dashed line at $\rho$=3 galaxies Mpc$^{-2}$ shows the threshold to separate group from field galaxies. 73 $\%$
 of all field galaxies are found at densities below this limit and 90$\%$ of all group member galaxies are above that.}
\label{density}
\end{figure}

Figure \ref{density} shows the histogram of the density distribution obtained with our method for the whole galaxy population considered in this work (black histogram) and for the galaxies identified as groups spectroscopic members (red histogram), as described in the previous section. As confirmed by the simulation described above, galaxies in massive halos are not located on the bottom of the density distribution. Thus, the density indicator can be used to efficiently isolate galaxies that are hosted by low mass halos. We use this histogram to define the density cut for defining the ``field'' galaxy sample, that is galaxies that should be hosted by DM halos of masses below $10^{12}$ $M_{\odot}$ according to our simulations.  The green dashed line at $\rho$=3 galaxies Mpc$^{-2}$ in Figure~\ref{density} shows the threshold to separate group from field galaxies. Indeed, 90$\%$ of all group member galaxies are above this limit. We do this exercise separately for the two redshift bins considered in our work. This leads to a sample of 4987 field galaxies in low redshift bin and 6063 field galaxies at high redshift bin.

Similarly to \cite{Ziparo2014}, we define a third environmental class of galaxies identified by density values similar to the ones of group galaxy members but not associated to any X-ray extended emission observed in the X-ray surveys considered in this work. These galaxies have density above the $\rho$=3 galaxies Mpc$^{-2}$ threshold and do not lie in the sky region defined by detected X-ray extended emissions. They likely belong to filaments, sheet like structures or to groups at lower mass with respect to the mass limit imposed by the CDFS, CDFN, AEGIS and COSMOS X-ray detection limits. We define this class of objects as ``filament-like'' galaxies. With this approach, we define a sample of 1246 ``filament like'' galaxies in low redshift bin and 2320 in high redshift bin. This additional class of objects will be used also to check whether the relative vicinity of galaxies can affect galaxy properties as suggested by \cite{Peng2010, Peng2012} or, instead, the membership to a massive halo is a key ingredient in the galaxy evolution.

 
  \subsection{Morphology}
\label{morphology}
We link the location of galaxies in the SFR-stellar mass plane also to the galaxy morphology. The galaxy morphological information are taken from the Advanced Camera for Surveys General Catalog (ACS-GC), which is a photometric and morphological database based on publicly available data obtained with the Advanced Camera for Surveys (ACS) instrument on board of the Hubble Space Telescope (\citealt{Griffith2012}). This catalog contains $\sim$370,000 galaxies observed in the EGS, COSMOS, GEMS and GOODS surveys. Briefly, \cite{Griffith2012} use GALFIT to measure the structural parameters of each galaxy by modeling each source with a single S\'ersic profile (\citealt{Sersic1986}; see \citealt{Graham2005} for the mathematical relationship) together with a model for the sky. By cross-matching our galaxy catalogs of EGS, COSMOS and the GOODS fields with the ACS-GC catalog, we find morphology information for 80\% of our galaxy sample. We use the structural measurements based on the ACS F850LP imaging in GOODS-South , based on the ACS F775W imaging in GOODS-North and on the ACS F814W imaging in COSMOS and AEGIS.

\section{Results}

\subsection{The non-linearity and dispersion of the SF galaxy Main Sequence}
\label{nonlinear}
\begin{figure*}
 \centering
 \includegraphics[width=16cm]{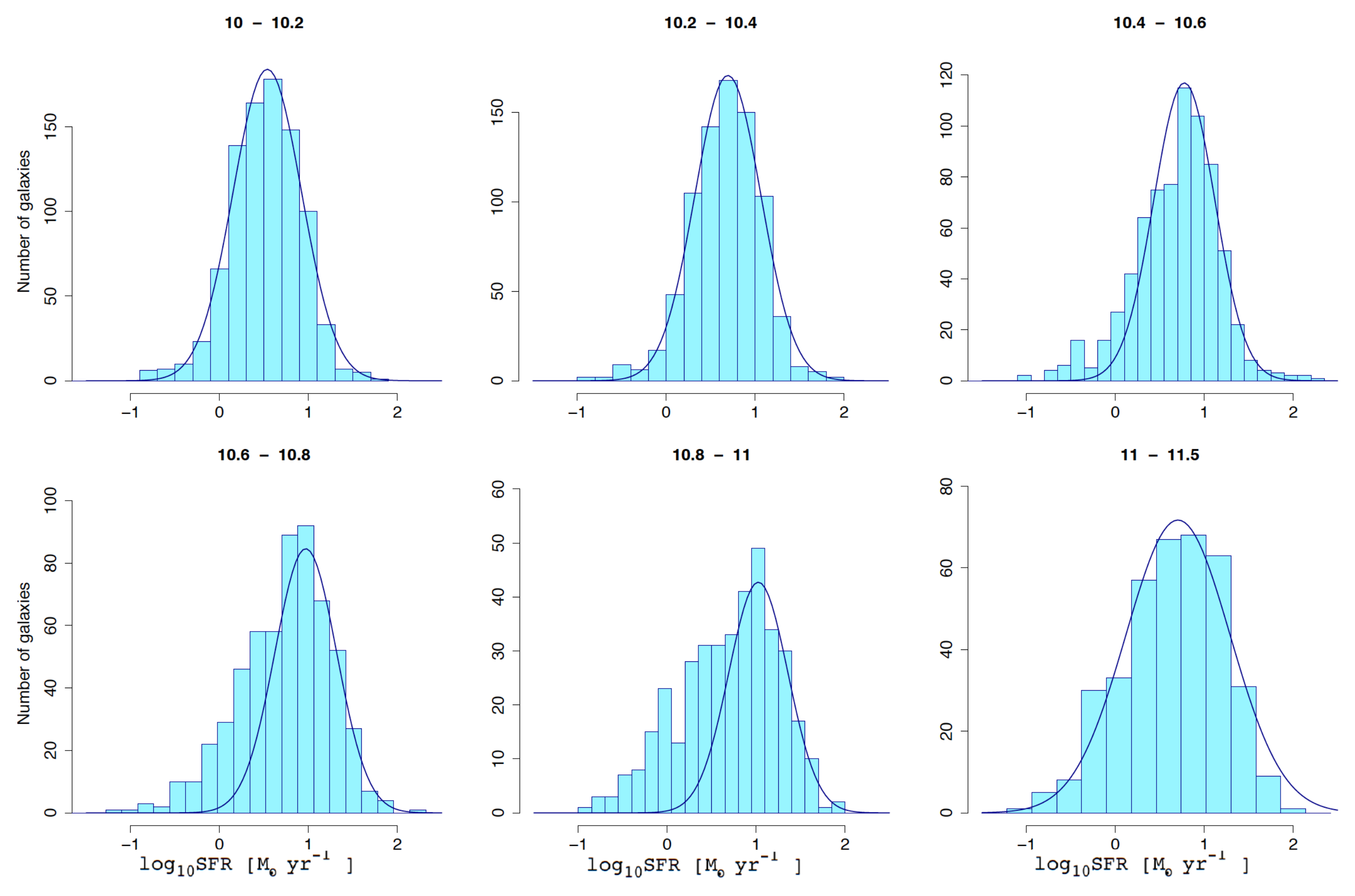}

\caption[Distribution of SFR of all IR detected galaxies at 0.15 $<z<$ 0.5 in different stellar mass bins.]{Distribution of SFR of all IR detected galaxies at 0.15 $<z<$ 0.5 in different stellar mass bins.}
\label{SFRM-M-lowz}
\end{figure*}

\begin{figure*}
 \centering
 \includegraphics[width=16cm]{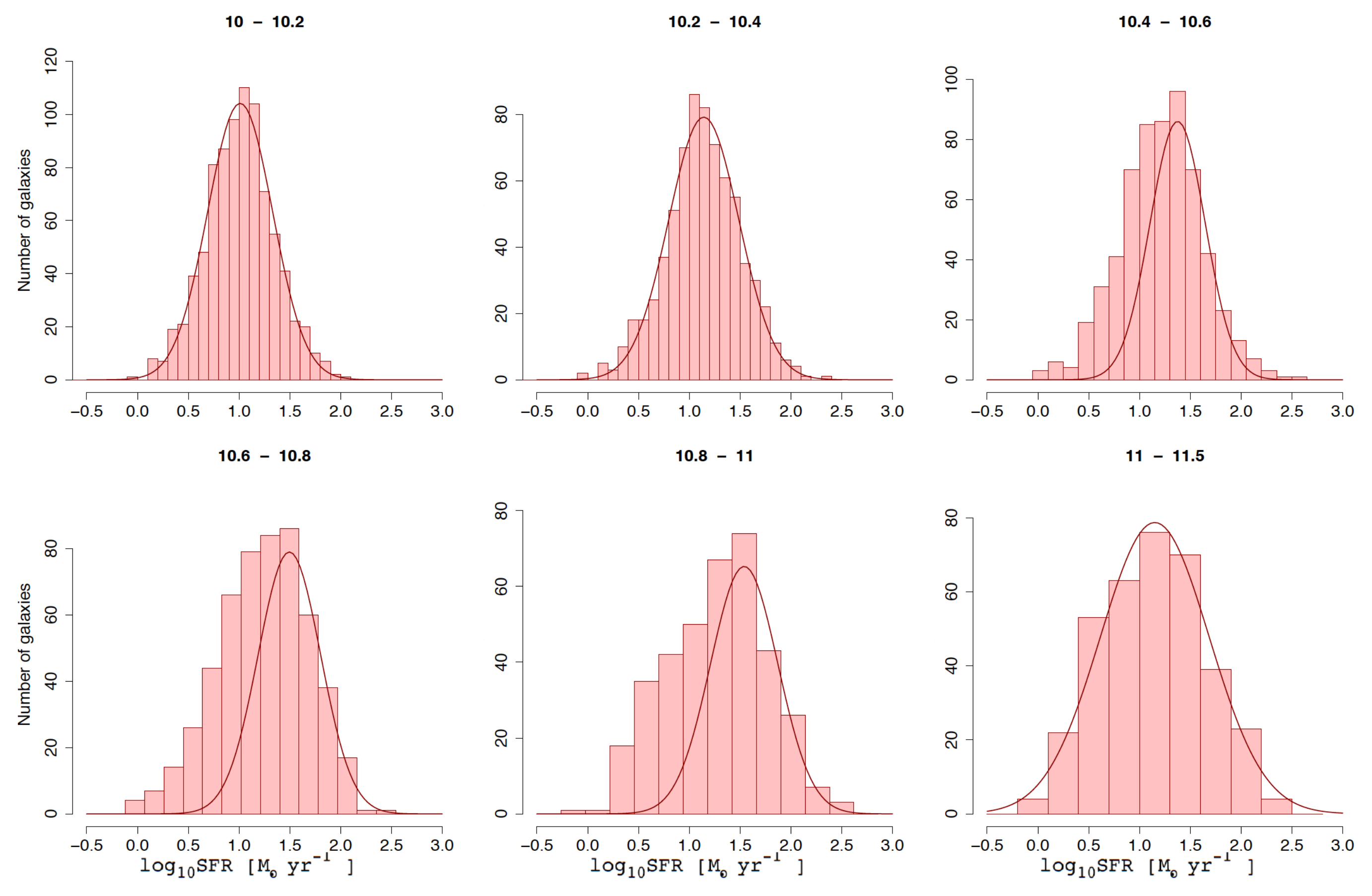}

 \caption[Distribution of SFR of all IR detected galaxies at 0.5 $<z<$ 1.1 in different stellar mass bins.]{Distribution of SFR of all IR detected galaxies at 0.5 $<z<$ 1.1 in different stellar mass bins.}
\label{SFRM-M-highz}
\end{figure*}

Before studying the location of galaxies in the SFR-stellar mass plane as a function of the environment, it is necessary to define the galaxy MS in our two redshift bins and estimate its dispersion. In doing this, we try to avoid any selection bias by applying the following procedure.  First, we investigate if the MS is a linear relation (i.e., with a slope of unity) or if there is some level of non-linearity (deviation from a slope of 1, and possible curvature) at the high stellar masses as suggested by \cite{Whitaker2012, Whitaker2014}. As a starting point, we look at the distribution of SFR of the IR detected galaxies in the photometric redshift sample to avoid any selection bias which can be induced by spectroscopic selection function. Of course $z_{phot}$ is replaced by the spectroscopic redshift whenever this is available. Figure \ref{SFRM-M-lowz} and \ref{SFRM-M-highz} show SFR$_{IR}$ distribution
in different stellar mass bins for low and high redshift samples, respectively. The Gaussian curves are fitted to the right side of the peaks of the distributions and their mirrors in the left sides. The dispersions of the Gaussians are about 0.27-0.3 dex , consistent with the values reported by many previous studies (\citealt{Daddi2007, Elbaz2007,Peng2010}). However, these figures demonstrate that a Gaussian is not a good fit to the MS at least above 10$^{10.4}$ $M_{\odot}$ where we are absolute complete in IR data. 

\begin{figure*}
 \centering
 \includegraphics[width=1\textwidth]{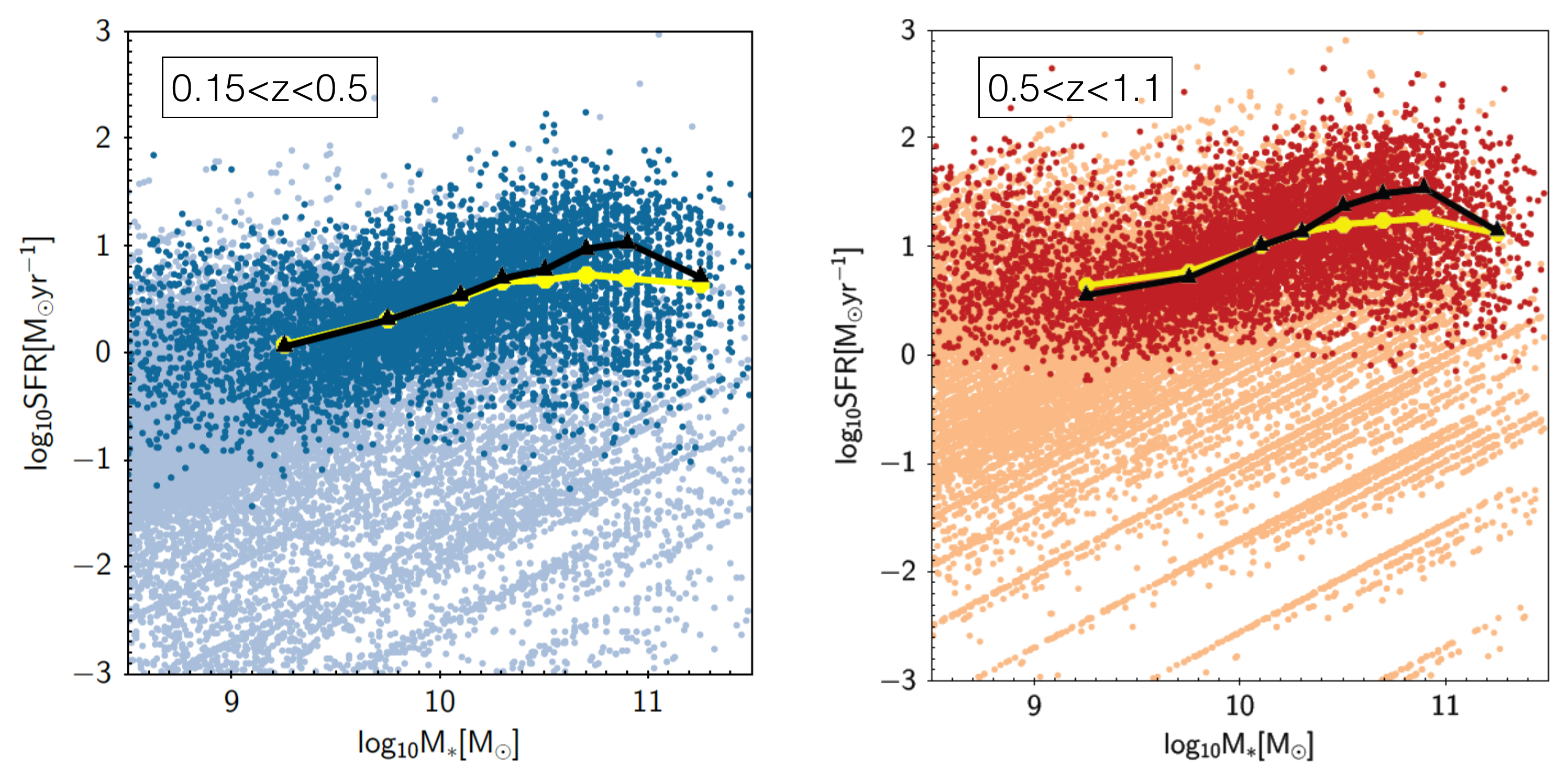}

  \caption[SFR vs. $M_*$ for galaxies for IR galaxies]{\textit{left panel} :SFR vs. $M_*$ for galaxies in 0.15 $<z<$ 0.5.
 The dark blue points show the IR detected galaxies and the light blue points show galaxies with SFR$_{SED}$. The black connected triangles show the peak of distribution of SFR in different mass bins for IR detected galaxies and the yellow connected circles show the mean of SFR for IR detected galaxies.
\textit{right panel}: Same as left panel for galaxies in 0.5 $<z<$ 1.1.}
\label{SFRM}
\end{figure*}

\begin{figure}
\centering
\includegraphics[width=0.4\textwidth]{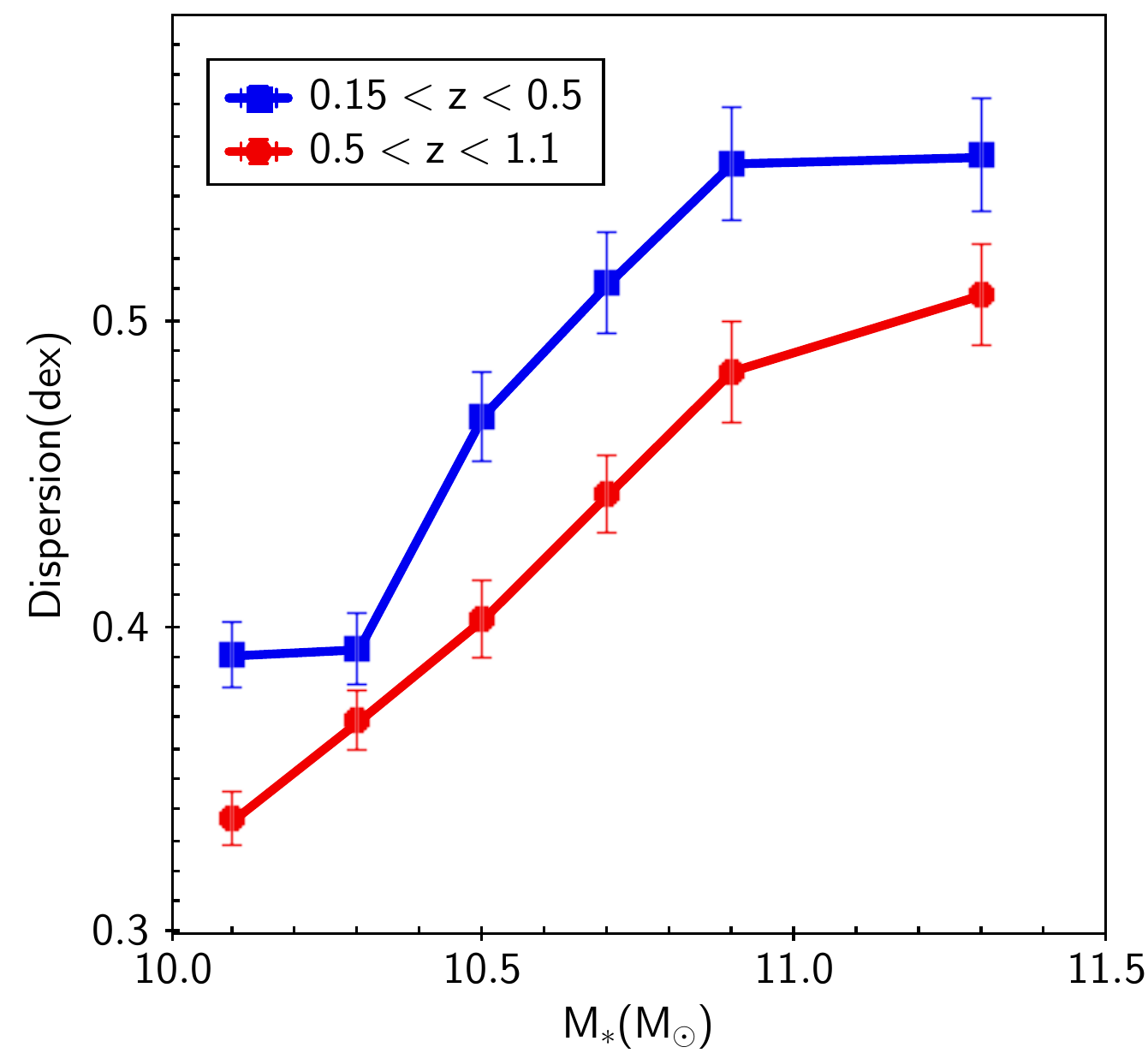} 
\caption[Dispersion around the MS location as a function of the galaxy stellar mass]{Dispersion around the MS location as a function of the galaxy stellar mass for high redshift bin (in red circles) and low redshift bin (in blue squares). Errors on the dispersion are calculated from bootstrapping.}
\label{summary}
\end{figure}

\begin{figure}
 \centering
 \includegraphics[width=0.45\textwidth]{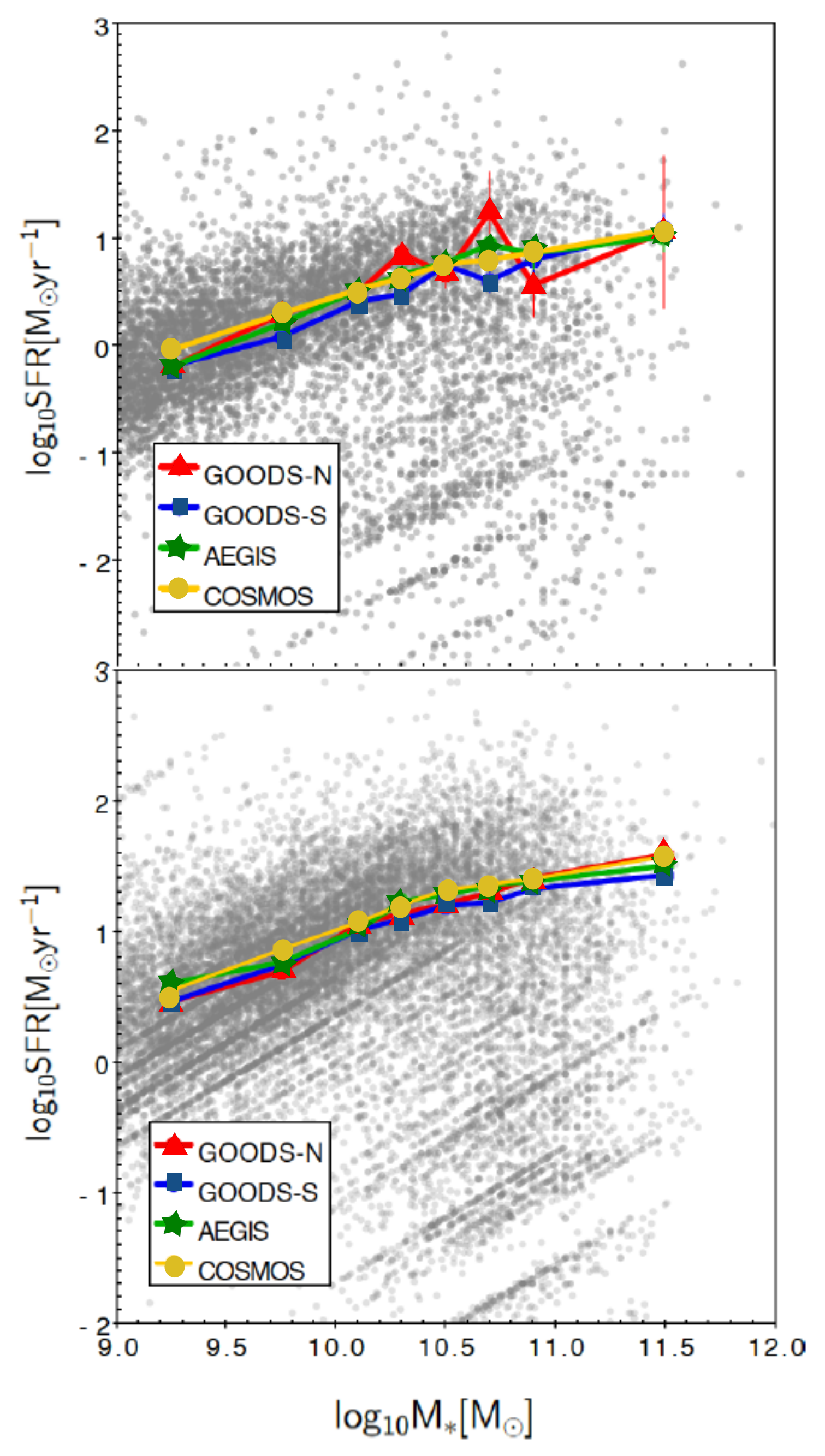}

  \caption[SFR vs. $M_*$ for galaxies for IR galaxies]{\textit{upper panel} :SFR vs. $M_*$ for galaxies in 0.15 $<z<$ 0.5. The colorful connected circles show the mean of SFR for star forming galaxies in different fields.
 \textit{lower panel}: Same as upper panel for galaxies in 0.5 $<z<$ 1.1.}
\label{diff_field}
\end{figure}

We define a reference MS relation. Since the MS is well studied in the literature (\citealt{Elbaz2007} for 0.8 $<z<$ 1.2, \citealt{Noeske2007a} for 0.2 $<z<$ 0.7 and \citealt{Peng2010} for 0.02 $<z<$ 0.085), we interpolate existing relations to retrieve the MS in the range of redshift used in this work.
\begin{equation}
logSFR=-7.9+0.82\times logM_*\\0.15<z<0.5
\end{equation}
\begin{equation}
logSFR=-7.5+0.83\times logM_*  \\0.5<z<1.1
\end{equation}
 We consider those galaxies within $\pm$ 1 dex of these interpolated main sequences as normal star-forming galaxies (consistent with 3$\times$ dispersion of MS reported in the literature). We use this criteria to choose the star-forming galaxies for our analysis in the rest of this paper. Fig. \ref{SFRM} shows the location of the mean and peak of SFR of the IR detected star-forming galaxies in the SFR-stellar mass plane in the two redshift bins. The peaks of the distributions are consistent with the interpolated MS. Below 10$^{10.4}$ $M_{\odot}$, the mean of SFR is highly consistent with the peak of distribution of SFR$_{IR}$. Above this threshold, the higher the stellar mass, leads to the larger deviation of the mean from the peak towards lower values of SFRs. This indicates a flattening of the MS at high stellar masses towards low SFR consistent with the findings of \cite{Whitaker2012}. This is actually the mass threshold where a tail appears in the SFR distribution (Fig. \ref{SFRM-M-lowz} and \ref{SFRM-M-highz}) and we don't have any more a Gaussian distribution for SFR of galaxies. Furthermore, we look at the standard deviation of the SFR in each stellar mass bin as a function of the galaxy stellar mass (Fig. \ref{summary}). Above 10$^{10.4}$ $M_{\odot}$, we observe an increase of the standard deviation around the mean of MS. The scatter of the MS is about 0.3-0.4 dex below 10$^{10.4}$ $M_{\odot}$, consistent with the values reported by many previous studies (\citealt{Daddi2007, Elbaz2007,Peng2010}). It increases to 0.5-0.6 dex at higher masses. Moreover,  the standard deviation seems to be larger at low redshift with respect to the high redshift MS, as shown in Fig. \ref{summary}.  In order to check for possible biases, we also investigate the MS in different fields, separately. Figure \ref{diff_field} demonstrates that different fields do describe a consistent star-formaing sequence. The most discrepancy is in the low redshift sample for GOODS-S and GOODS-N fields. The reason is the low number of galaxies in each bin in these fields at low redshift. However, they are also consistent with other fields within their error bars. 
  
In several works in the literature the MS of star forming galaxies is expressed through a linear relation with slope consistent to 1 (\citealt{Elbaz2007} for 0.8 $<z<$ 1.2, \citealt{Noeske2007a} for 0.2 $<z<$ 0.7 and \citealt{Peng2010} for 0.02 $<z<$ 0.085). \cite{Peng2010}, in particular, show that the MS of blue star forming galaxies selected from the SDSS spectroscopic catalog is linear up to very high masses and  its slope and dispersion is independent from the environment. However, the selection of only blue galaxies as a way to isolate the bulk of the star forming galaxies might affect their results. Indeed, \cite{Weinmann2006} shows that 20\% of the galaxies hosted by massive halos such as groups and clusters show red colors but a level of star formation activity similar to the blue active galaxy population. In addition, \cite{Brinchmann2004} show that, when  all galaxies are considered, the MS of the local Universe is well identified at stellar masses below $10^{10-10.5}$ $M_{\odot}$ but it breaks down at higher masses. More recently, \cite{Whitaker2014} show that the deviation from a linear relation of the MS at high stellar masses is evident up to redshift $\sim 2$. This would be consistent also with the relatively flat slope of the MS found by \cite{Rodighiero2010} up to $z\sim$ 2.5, obtained by stacking analysis in the Herschel PACS data.

\begin{figure*}
 \centering
 \includegraphics[width=8.5cm]{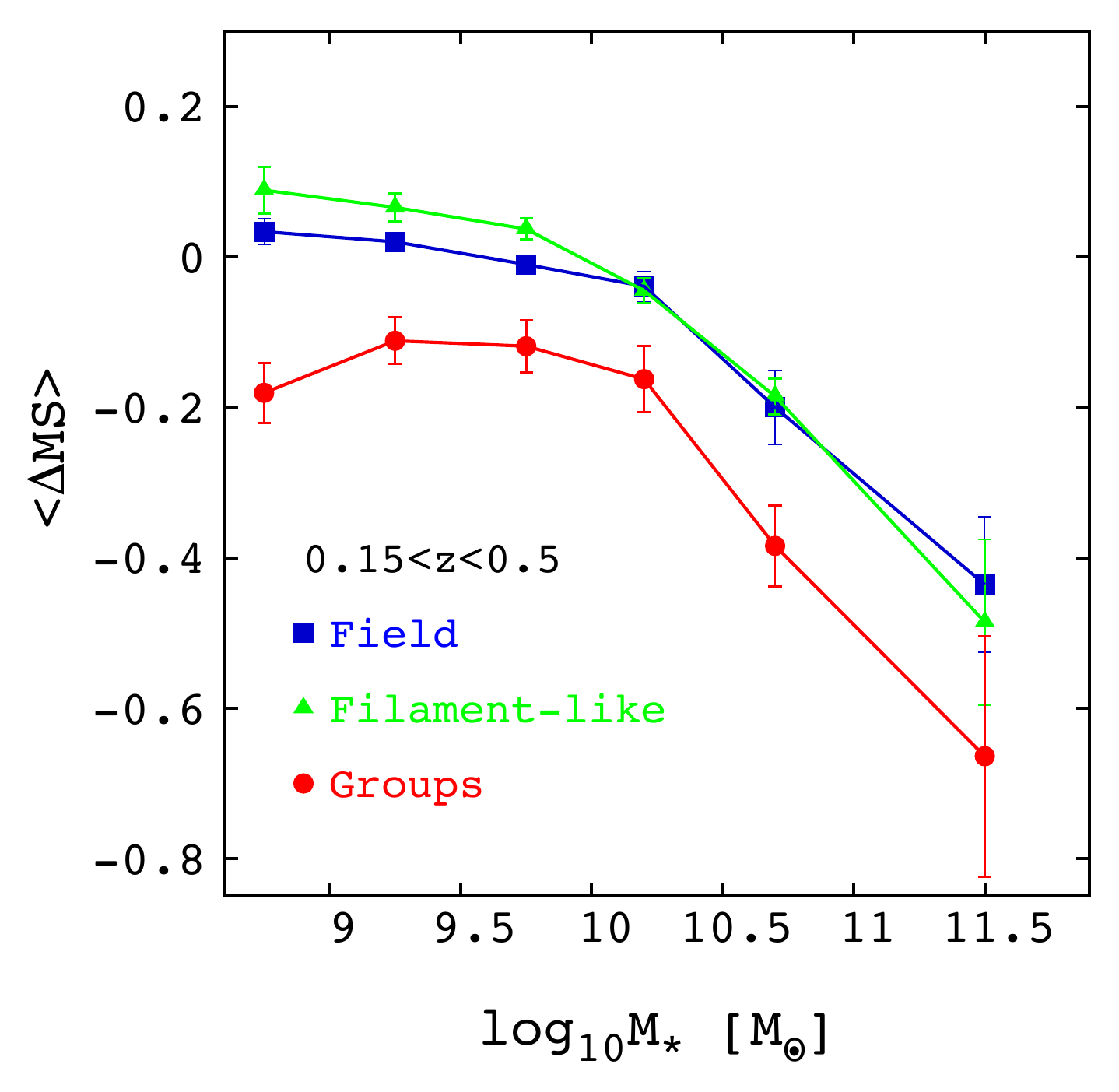}
\includegraphics[width=8.5cm]{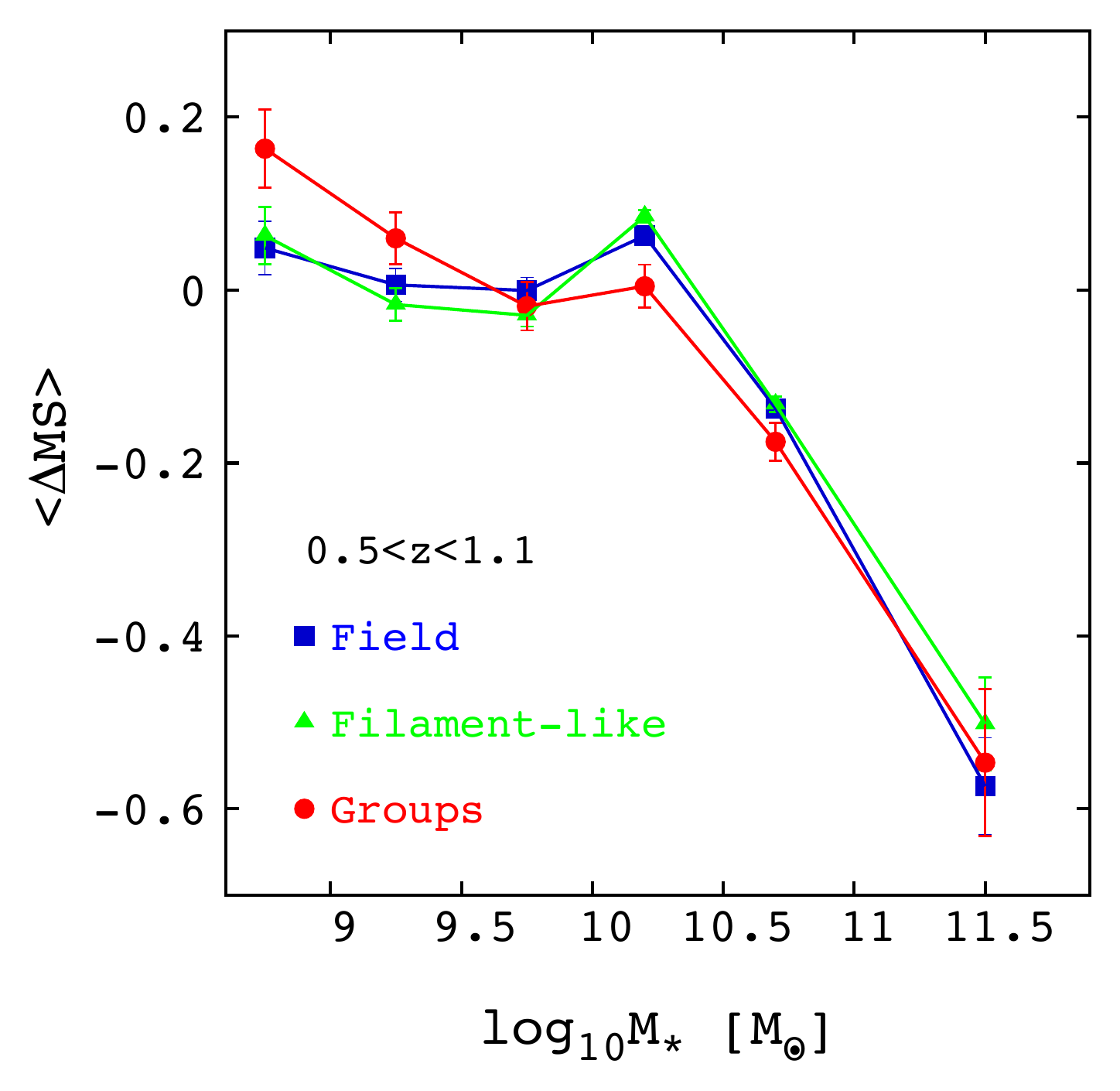}
\caption{The MS offset for field, ``filament-like" and group star forming galaxies as a function of stellar masses. $<\Delta$MS$>$ represents the mean of the residuals with respect to predicted MS in different stellar mass bins for low redshift bin (left panel) and high redshift bin (right panel).  Errors are based on standard errors (see Appendix \ref{App:AppendixA}).  }
\label{deltaM}
\end{figure*} 

\begin{figure*}
 \centering
 \includegraphics[width=17cm]{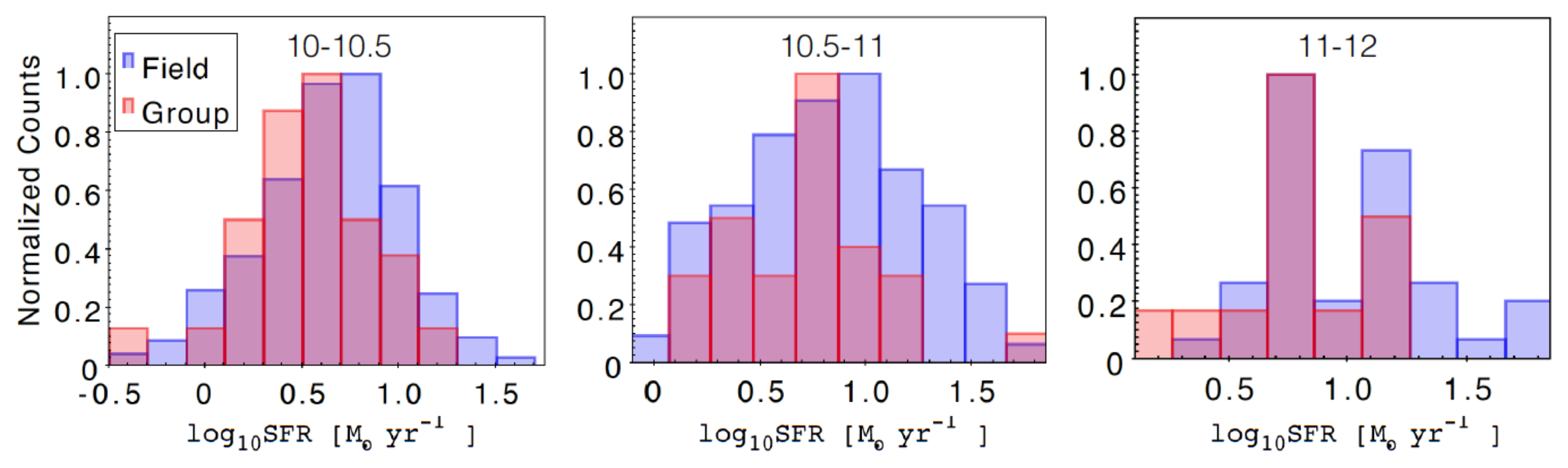}
\caption{The distribution of star formation of IR detected star forming galaxies for the field (in blue) and the group galaxies (in red) in different stellar mass bins at low redshift.}
\label{deltaMIR}
\end{figure*}

\subsection{The role of environment in shaping the Main Sequence}

To investigate if the environment plays any role in shaping the MS, we analyse the deviation of the MS from the linear relation in the three environments defined in Section \ref{env}. This is done to check if the environment can be a cause either of the flattening of the MS or of the larger dispersion at high stellar masses. Figure \ref{deltaM} shows the main sequence offset with respect to the reference linear relation ($\Delta$MS) as a function of stellar masses for field galaxies (blue points), ``filament-like'' galaxies (green points) and group galaxies (red points) in the low redshift bin (left panel) and in the high redshift bin (right panel). The analysis of these two plots lead to the following conclusions:

\begin{itemize}
\item [-]the deviation of the MS from the linear relation at stellar masses above $10^{10.4}$ $M_{\odot}$ exists in all environments
\item [-]this deviation is in place already at $z\sim$ 1.1
\item [-]at low redshift, group galaxies show a much more significant departure from the mean MS field at any stellar mass with respect to field and  ``filament-like" galaxies
\item [-]at higher redshifts, groups member galaxies do not deviate from the mean relation and their MS coincides with the MS of the other two environments
\item [-]field (isolated) and ``filament-like'' galaxies MS are perfectly consistent at any redshift. 
\end{itemize}

The last evidence shows that the relative vicinity of galaxies as expressed by the density field is not playing an important role in affecting and/or regulating the galaxy SF activity. This, in addition to the blue galaxy selection,  forms the likely explanation why \cite{Peng2010} did not observe any difference between the MS location of galaxies at different densities.

Figure \ref{deltaM} indicates clearly that the evolution of the star formation activity in galaxies does not depend only on the galaxy stellar mass through galaxy internal process (e.g. AGN feedback) but it must be regulated by also the environment in particular at the high stellar masses. Figure \ref{deltaMIR} shows the SFR distributions of the IR detected galaxies in group and field environments at low redshift. It illustrates that group galaxies are mostly populating the fading tail of the SFR distribution in Fig. \ref{SFRM-M-lowz} . Our results confirm the first indication of \cite{Ziparo2013} that group galaxies evolve in a much faster way with respect to galaxies in lower mass halos in terms of quenching of the SF activity. Thus, the membership to a massive halos and the effect of all physical processes in place in such halos are likely to be responsible for the decrease of the SF activity in groups since $z\sim$ 1. Figure \ref{deltaM} also explains the increase of the dispersion of the MS in the low redshift bin, as a function of the stellar mass shown in the Figure \ref{summary}. Indeed, the deviation of group galaxies towards a flatter  and lower MS with respect to the mean relation has the effect of a broadening of the mean MS.

 \subsection{The role of morphology in the shape and dispersion of Main Sequence}

\begin{figure*}
 \centering
 \includegraphics[width=0.48\textwidth]{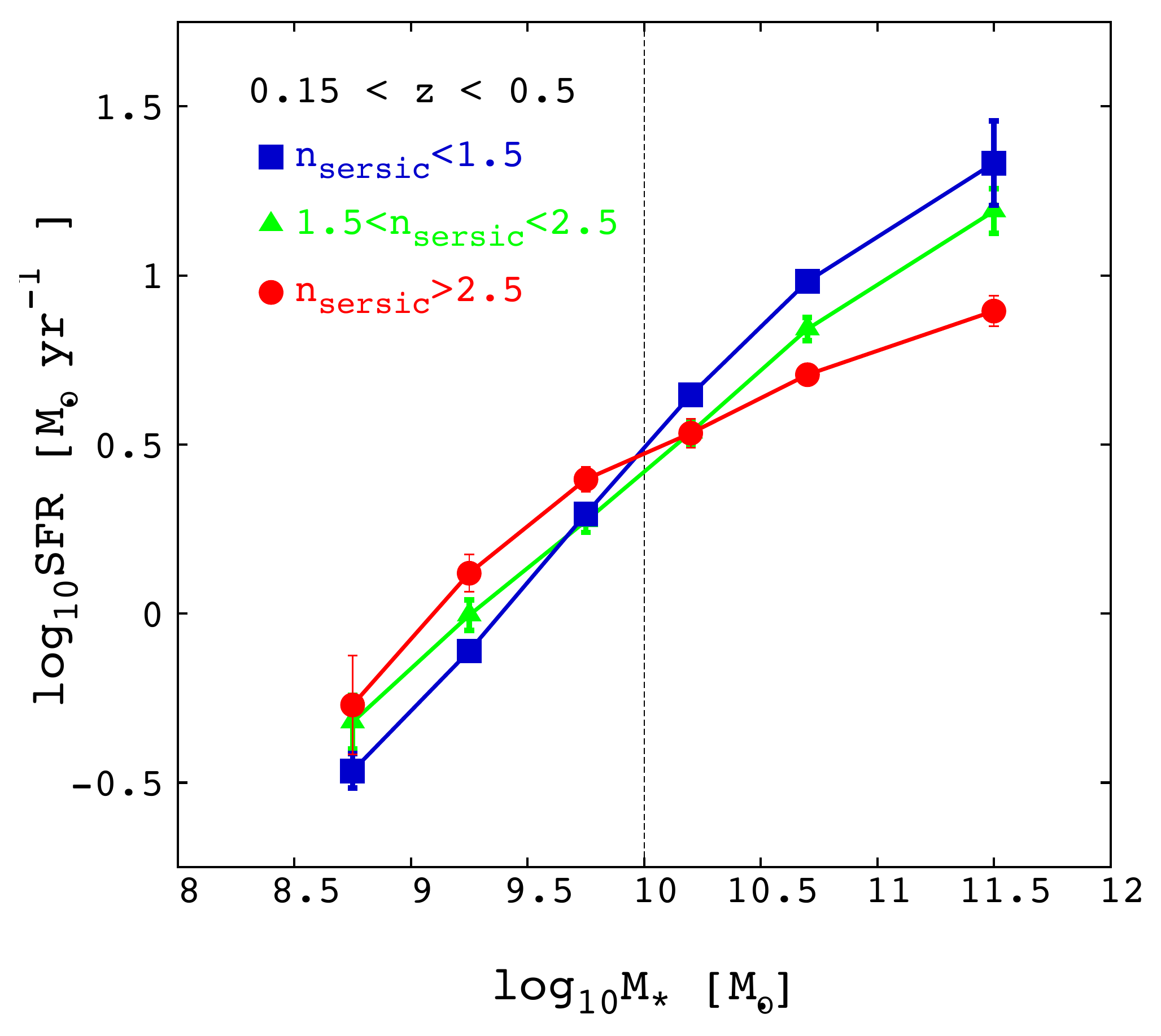}
 \includegraphics[width=0.48\textwidth]{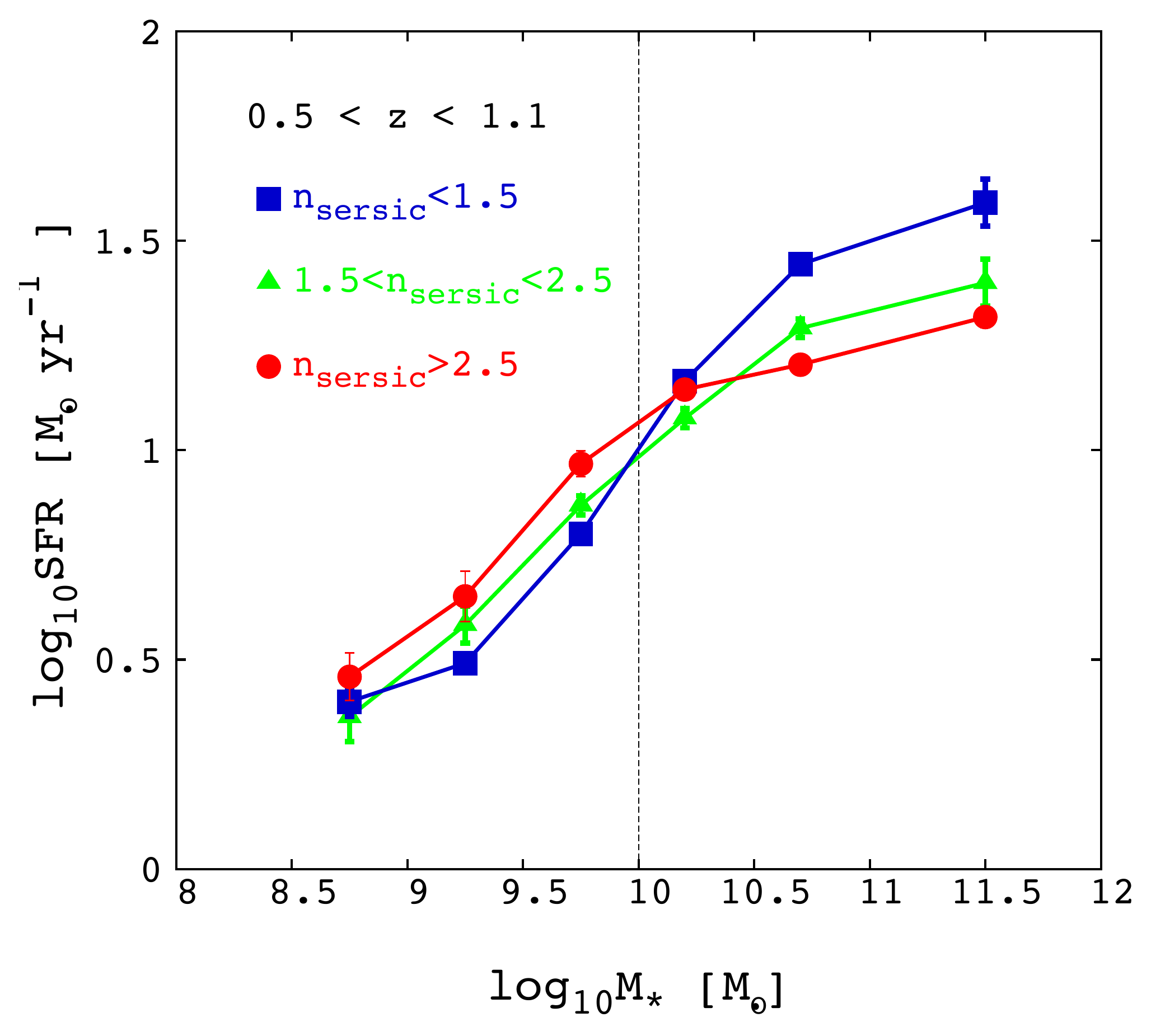}

  \caption{The mean of dependence of log SFR on stellar masses of MS galaxies in three different S\'ersic index ranges for the low (left panel) and high redshift bin (right panel).}
\label{morphology-all}
\end{figure*}

\begin{figure}
\includegraphics[width=0.48\textwidth]{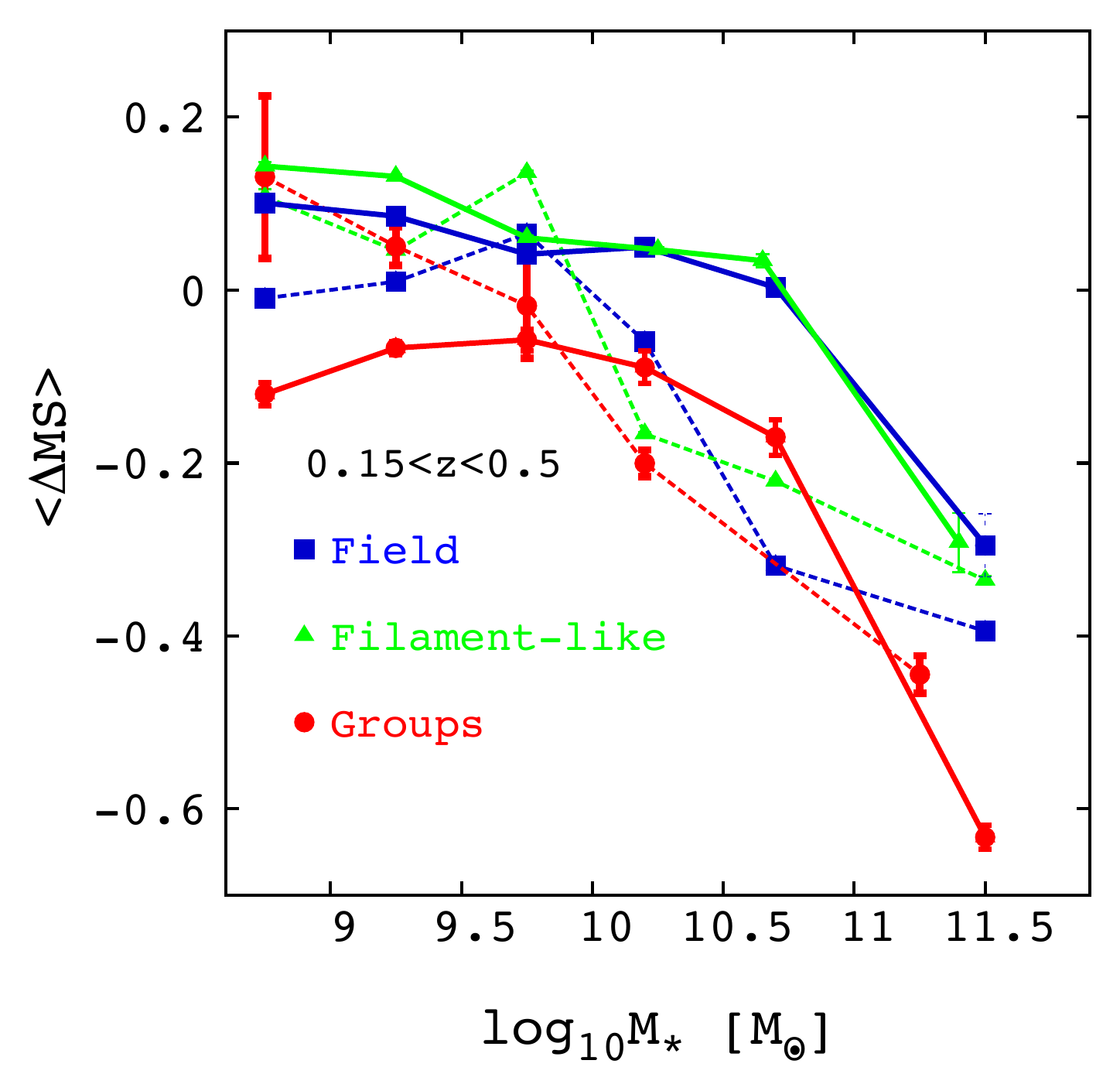}
\caption{The MS offset for field, ``filament-like" and group star forming galaxies as a function of stellar masses in the low redshift bin for disk dominated (n$_{sersic} < 2$ with solid lines) and bulge dominated galaxies (n$_{sersic} > 2$ with dashed lines).\label{deltaM-sersic-low}}
\end{figure}

\begin{figure}
\includegraphics[width=0.48\textwidth]{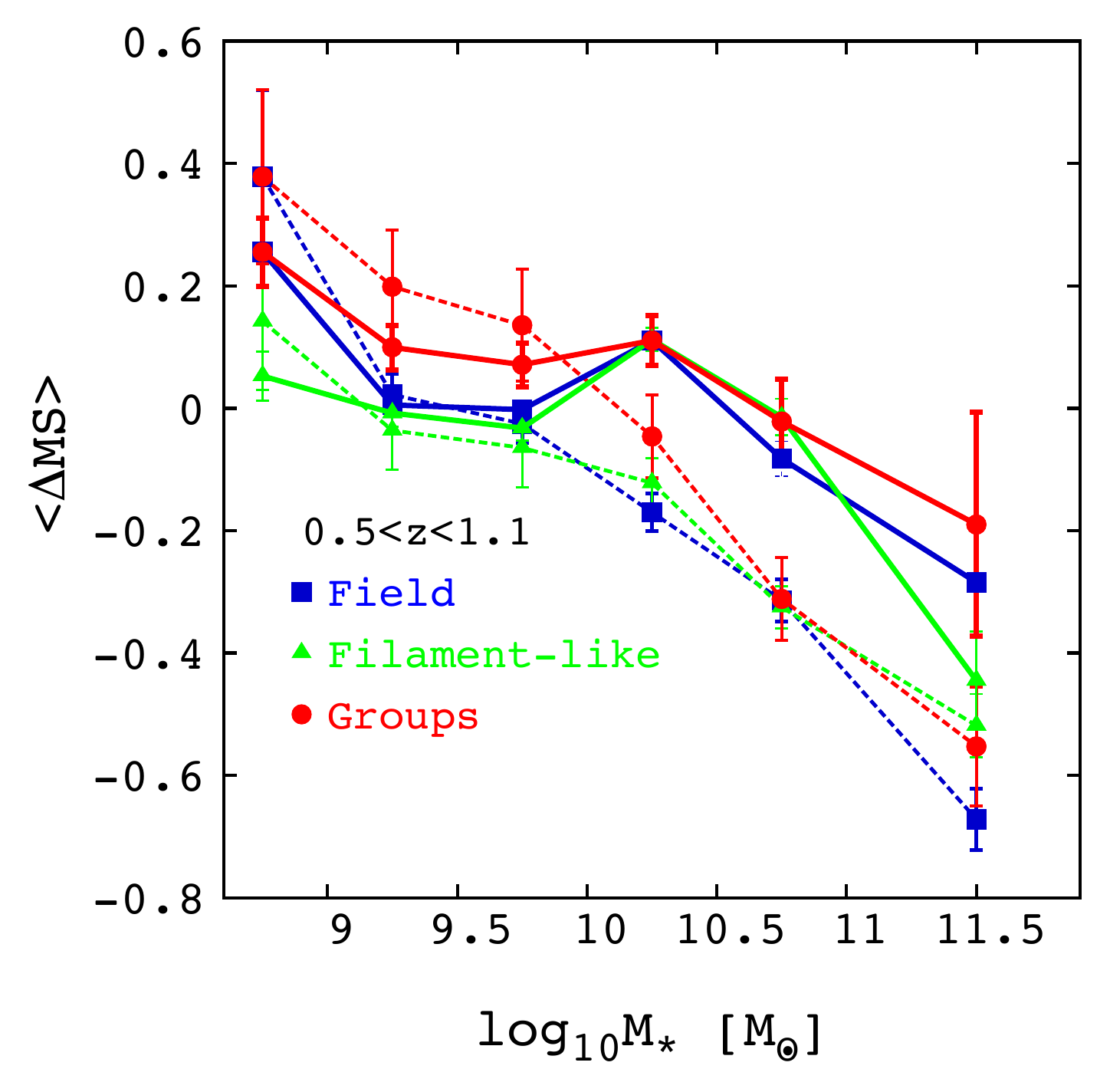}
\caption{The MS offset for field, ``filament-like" and group star forming galaxies as a function of stellar masses in the high redshift bin for disk dominated (n$_{sersic} < 2$ with solid lines) and bulge dominated galaxies (n$_{sersic} > 2$ with dashed lines).\label{deltaM-sersic-high}}
\end{figure}

To make a step further, we investigate also the relation of morphology, SFR and environment in the SFR-stellar mass plane. This is done to understand if the quenching of the SFR in group galaxies is also associated to a morphological transformation. For this purpose we study the location of the MS in different environments as a function of the morphology. We identify three classes of galaxies on the basis of the  S\'ersic index value (see Section \ref{morphology}): galaxies with S\'ersic index below 1.5 are classified as disk dominated, galaxies with values between 1.5 and 2.5 are classified as intermediate type galaxies, and galaxies with S\'ersic index above 2.5 are classified as bulge dominated systems. Consistently with  \cite{Wuyts2011}, above 10$^{10}$ $M_{\odot}$ galaxies are located across the MS as a function of the morphology. Disk dominated galaxies populate the upper envelope and intermediate type galaxies are just above the bulge dominated systems located on the lower envelope of the MS in any redshift bin, as shown in Fig.  \ref{morphology-all}. We point out that the flattening of the MS above the stellar mass threshold of $10^{10.4}$$ M_{\odot}$ seems to be more evident for the bulge dominated galaxies while disk dominated galaxies tend to show a more linear relation. This result would indicate that the flattening of the MS is more related to the morphological evolution of galaxies rather than the effect of the environment. In addition, the displacement of bulge dominated galaxies at lower SFR in the high stellar mass regime could contribute to the larger scatter of the relation in the high stellar mass regime.

We further study the location of the MS as a function of morphology and environment in Fig. \ref{deltaM-sersic-low} and \ref{deltaM-sersic-high}. These figures show the mean $\Delta$MS per stellar mass bin of each class of galaxies with respect to the reference MS linear relation in the low and high redshift bins. In order to have enough statistics, we divide galaxies only in two morphological classes: disk dominated galaxies (S\'ersic index $< 2$) and bulge dominated ones (S\'ersic index $> 2$). Figure \ref{deltaM-sersic-high} shows that, at high redshifts, in all environments, disk dominated  and bulge dominated galaxies are confirmed to occupy the upper and the lower envelope of the MS, respectively. At low redshift, in the group regime, instead, at high stellar masses also the disk dominated systems are located on the lower envelope of the relation as the bulge dominated counterpart ( Fig. \ref{deltaM-sersic-low}). This illustrates that the suppression of star formation in disk dominated galaxies causes the deviation of the group star forming galaxies from other environments towards lower SFR. This also indicates that a quenching of the SFR is happening before any morphological transformation in the group regime.  As a further step, we check the fraction of star forming  bulge dominated galaxies to the total star forming galaxies as a function of environment. To investigate the differences among the different environmental classes,  we calculate the fraction in several mass bins. This is done to take into account the stellar mass dependence of the flattening of the MS and the differential role of the environment at different mass scales.
\begin{figure*}
 \centering
 \includegraphics[width=8cm]{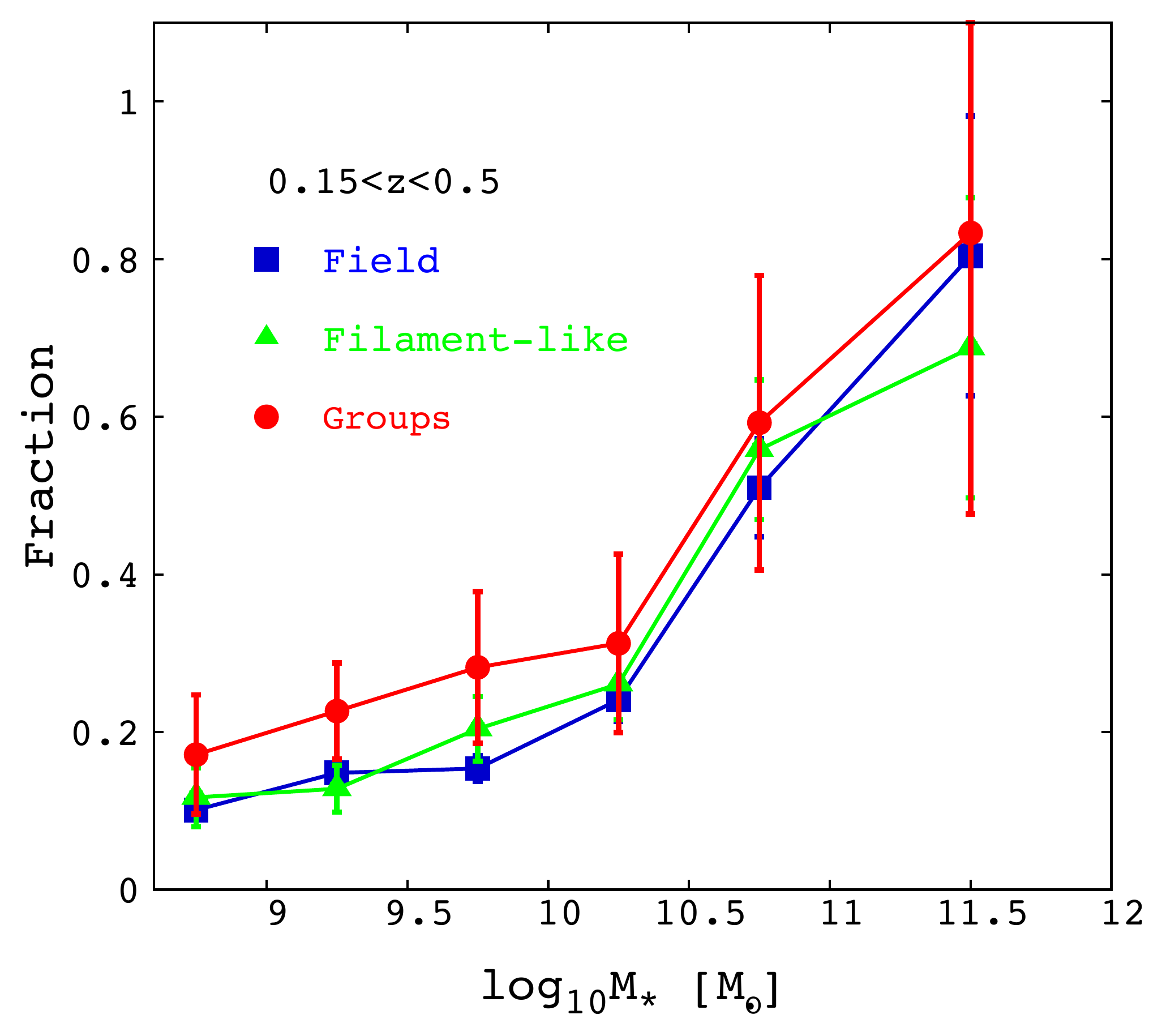}
\includegraphics[width=8cm]{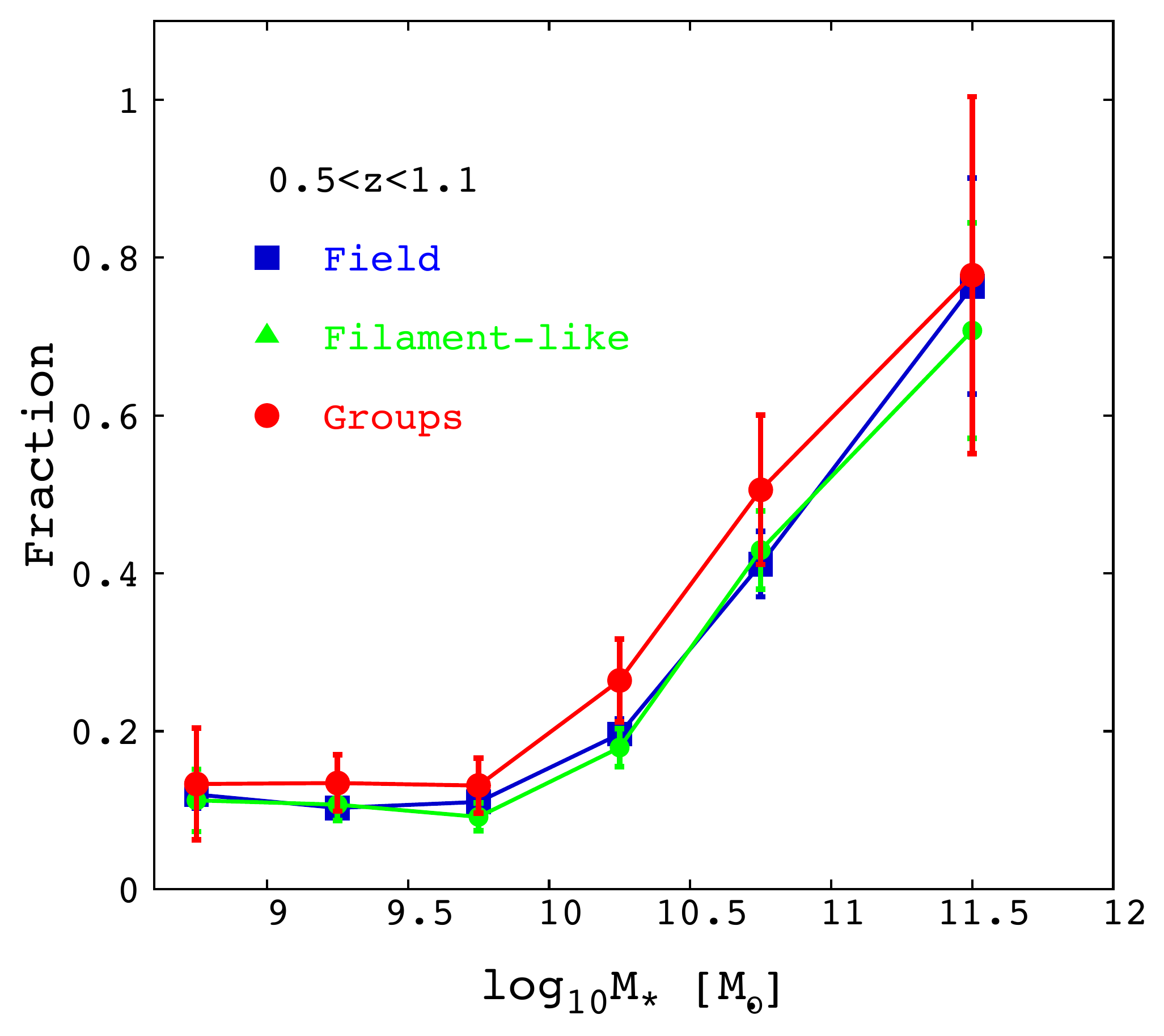}

  \caption[Mean of the S\'ersic index n as a function of the stellar mass]{\textit{left panel} : Fraction of MS galaxies with S\'ersic index n$>$ 2 to total MS galaxies as a function of the stellar mass in the low redshift bin in three environmental classes: groups (red points and line), ``filament-like'' galaxies (green points and line) and field galaxies (black points and line). \textit{right panel} : same as the left panel for the high redshift bin.}
\label{morpho}
\end{figure*}
Figure \ref{morpho} shows that the S\'ersic index distribution for MS galaxies is quite similar in all environmental classes both at
 low (left panel) and high (right panel) redshift.\\
 Figure \ref{morpho} in addition to Figure \ref{deltaM-sersic-low} suggest that the SF quenching observed in group
 galaxies in Figure \ref{deltaM}  is not associated to systematic morphological transformation. We stress that this is not at odds
 with the well known ``morphology-density'' relation. Indeed, this more general relation reflects the differential morphological
 type distribution of the entire galaxy population without distinction between galaxies on and off the Main Sequence. In this particular
 analysis, we consider only MS galaxies to show that the departure of group galaxies from the MS is not related to a morphological
 transformation. Instead, Figure \ref{morpho} shows clearly a much stronger dependence of the S\'ersic index distribution on the stellar
 mass at low and high redshift. While below $10^{10.4}$ $M_{\odot}$, MS galaxies tend to be late type, above this threshold the
 morphological type distribution is clearly dominated by intermediate type (1.5$<$S\'ersic index$<$2.5) and early type (S\'ersic index$>$2.5) galaxies. This fact in addition to the shallower slope for bulge dominated MS galaxies explain why we measure a shallower slope for MS galaxies.



\section{Discussion}

In this paper, we present the properties of the star formation Main Sequence and its connection to morphology and host environment of galaxies since $z\sim$ 1.1. Both the dispersion and the slope of the MS unravel fundamental information about the efficiency and stochasticity of the processes which regulate the star formation.  We see that the distribution of SFR for IR detected galaxies is not a Gaussian distribution till 10$^{11}$ $M_{\odot}$. The distribution of SFR at high stellar masses is skewed towards lower SFR at low and high redshift.  However, the peak of distribution is consistent with a linear relation as found in previous studies. Above 10$^{11}$ $M_{\odot}$, either there is no MS at all or there is a shift of of the peak towards low SFR. The existence of a long tail of the SFR distribution towards low SFR lead to an increase of the dispersion of the MS at higher masses.  Below 10$^{10.4-10.6}$ $M_{\odot}$ the dispersion is 0.3-0.4 dex, consistent with the value reported by many of previous studies (\citealt{Daddi2007, Elbaz2007, Peng2010}). Above this mass limit it increases up to 0.5-0.6 dex.

In agreement with our results \cite{Lee2015}, using far-infrared photometry from $Herschel$-PACS and SPIRE and $Spitzer$-MIPS 24 $\mu$m, find that the relationship between median SFR and $M_*$ follows a power-law at low stellar masses, and flattens to nearly constant SFR at high stellar masses at $z <$ 1.3.  \cite{Whitaker2014} also find that the slope of the MS is dependent on stellar mass. They find that MS has roughly constant slope of unity for $M_*$$< 10^{10.2}$ $M_{\odot}$ and a shallower slope for the high mass end spanning the redshift range 0.5 $<z<$ 2.5. In addition, \cite{Huang2012} using a sample of galaxies in the local universe from SDSS suggest a transition mass of 10$^{9.5}$ $M_\odot$ below which star formation scales differently with total stellar mass  (also found by \citealt{Salim2007}) with a steeper slope.  \cite{Rodighiero2010} also find a rather flat MS at any redshift on the basis of Herschel PACS data. \cite{Leja2015} find, by exploring the connection between the observed MS and stellar mass function, that the shallow slope for the MS can not be hold for the low masses. 

According to Figure \ref{morphology-all} and \ref{morpho} , bulge dominated galaxies are the main deriver of the flattening of the MS since galaxies with lower S\'ersic index ( disk dominated galaxies with n$_{Sersic}$$<$1.5) have a steeper MS. Furthermore, the presence of such bulge dominated galaxies in the lower envelope of the MS explains the increase in the width of MS for the high mass end.  \cite{Wuyts2011} find that across cosmic time, the typical S\'ersic index of galaxies is optimally described as a function of their position relative to the MS at the epoch of their observation. The correspondence between mass, SFR, and structure, as quantified by the S\'ersic index, is equivalent to the Hubble sequence. Based on a quite similar galaxy sample (using SDSS for nearby universe and COSMOS, UDS and GOODS fields for high redshift samples), \cite{Wuyts2011} see that such a sequence already existed at $z\sim$ 2; bulge-dominated morphologies go hand in hand with a more quiescent nature. In addition, they observe that late type galaxies (S\'ersic index $< 1.5$) follow a linear SFR-Mass relation. Indeed, we also observe that late type galaxies follow a linear relation (see Figure \ref{morphology-all}). Nevertheless, they do not represent the bulk of the Main Sequence galaxy population above $10^{10.4}$ $M_{\odot}$, but only the upper envelope.
 This means that, while a clear morphological sequence with a linear slope and, thus, a strong stellar mass dependence, is visible in the SFR-Stellar mass plane at any mass as shown by \cite{Wuyts2011}, the Main Sequence of star forming galaxies at high masses is much more poorly defined and it shows a much weaker dependence on the stellar mass. This MS  is consistent with a
 linear relation only in the low stellar mass range where it is dominated by late type galaxies. At high masses, late type galaxies are no longer the bulk of the MS population. There, morphological sequence and star forming galaxy sequence are no longer overlapping. In agreement with our results, \cite{Whitaker2012} also find by selecting blue galaxies, the slope of unity for MS will be reached consistent with \cite{Peng2010}.  Consistently with our results, \cite{Lang2014}, using a mass selected galaxies, above 10$^{10}$ $M_{\odot}$, observe a rising trends of the median S\'ersic index and B/T ratio of star forming galaxies with increasing stellar mass at $z\sim$ 1 and $z\sim$ 2. Moreover, using a sample of SDSS galaxies, \citealt{Abramson2014} find that renormalizing SFR by disk stellar mass reduces the $M_{*}-$dependence of SF efficiency by $\sim$ 0.25 dex per dex , reducing the slope by 0.25. They also suggest that the nonlinearity of MS may simply reflect the well-known increase in bulge mass-fractions with stellar mass. \\

Our study of the MS in different environments demonstrates that the environment does not seem to be the cause of the flattening of MS at high stellar masses, since the MS shows the same type of flattening in all environments. Moreover, at high redshift, group galaxies are perfectly on sequence as galaxies in the other environments. We only observe a significant decline of SFR of star forming group galaxies at low redshift. This is consistent with previous studies in the literature. Indeed, the environmental trends at fixed stellar mass seem to weaken at higher redshift (e.g. \citealt{Poggianti2008, Tasca2009,Cucciati2010, Iovino2010, Kovac2010, Ziparo2014}). More recently, \cite{Popesso2015a} by the analysis of the Infra-Red (IR) Luminosity function in groups find at $z\sim$ 1, LIRGs and ULIRGs have contributions about 70\% and 100\% respectively to group galaxies while these contributions reach to less than 10\% for the nearby groups. Consistently with these results,  \cite{Popesso2015b} , show that the cosmic star formation history declines faster in group sized halos. Indeed,  we also observe a faster evolution in the star formation activity of star forming galaxies inhabiting in group environments . We find this faster evolution is independent of stellar mass. Indeed, this faster evolution and decline of star formation activity in group environments acts in terms of increasing the scatter around the MS. In Figure \ref{summary} also there is evidences for our claim as we measure a larger width for the MS in low-z sample.
In addition, numerous previous studies of low redshift galaxies in the literature find that color and star formation rate are more strongly correlated with the environment than morphology (e.g. \citealt{Kauffmann2004,Blanton2005,Christlein2005,vandenBosch2008,Weinmann2009}).
 The implication of these studies is that the well-known correlation between morphology and environment is secondary
 to the correlation between environment and star formation rate. This is consistent with our findings. Indeed, we observe that among
 MS galaxies the distribution of the morphological type does not depend on the environment. However, at the same time group galaxies tend to be more quenched with respect to field and filament like galaxies. Our results imply that in mass quenching, there is a morphological transformation (bulge growth) preceding quenching, but in environmental quenching the quenching happens before any morphological transition (if the latter happens at all).  \\
 
Our results suggest that above a mass threshold ($\sim 10^{10.4}-10^{10.6}$ $M_{\odot}$), stellar mass, morphology and environment act together in driving the evolution of the SF activity towards lower level. The presence of a massive bulge could be responsible of the so called ``morphological quenching'' suggested by \citealt{Martig2009}. In this model, the presence of a dominant bulge stabilizes the gaseous disk against gravitational instabilities needed for star formation. This would be consistent with the results obtained by \citealt{Whitaker2015}, using a photometric mass-complete sample of galaxies at 0.5 $<$ z $<$ 2.5. They find that the scatter of the MS is related in part to galaxy structure. According to their results, they suggest a rapid build-up of bulges in massive galaxies at $z \sim 2$. At $z$ $<$ 1, the presence of older bulges within star-forming galaxies decreasing the slope and contributing significantly to the scatter of the MS. \\
 Another possibility may be the growth of the central supermassive black hole which is the primary quenching agent for massive galaxies and is tightly coupled with the growth of bulges through both merging and disk instabilities (see SAM of \cite{Somerville2008}). We note that this process is in place beyond $z\sim$ 1.1 (\citealt{Lang2014, Whitaker2014}). The environmental effects appear, instead, at lower redshifts. Our results suggest that the environmental process, responsible for the suppression of star formation activity, should be a slow one as also suggested by \cite{Popesso2015b} and in agreement with \cite{Wetzel2013},  \cite{Delucia2012}, \cite{Wheeler2014} and \cite{Simha2009}.

\section*{Acknowledgements}
 The authors acknowledge Alvio Renzini and Michael L. Balogh for their useful comments and discussion on the early draft. 
 
 PACS has been developed by a consortium of institutes led by MPE 
(Germany) and including UVIE (Austria); KUL, CSL, IMEC (Belgium); CEA, OAMP (France); MPIA (Germany); IFSI, OAP/AOT, OAA/CAISMI, LENS, SISSA 
(Italy); IAC (Spain). This development has been supported by the funding 
agencies BMVIT (Austria), ESA-PRODEX (Belgium), CEA/CNES (France),
DLR (Germany), ASI (Italy), and CICYT/MCYT (Spain).

This research has made use of NASA's Astrophysics Data System, of NED,
which is operated by JPL/Caltech, under contract with NASA, and of
SDSS, which has been funded by the Sloan Foundation, NSF, the US
Department of Energy, NASA, the Japanese Monbukagakusho, the Max
Planck Society, and the Higher Education Funding Council of England.
The SDSS is managed by the participating institutions
(www.sdss.org/collaboration/credits.html).

We gratefully acknowledge the contributions of the entire
COSMOS collaboration consisting of more than 100 scientists.
More information about the COSMOS survey is available at
http://www.astro.caltech.edu/∼cosmos.
This project has been supported by the DLR grant 50OR1013 to MPE.\\
\\

\bibliographystyle{mnras}
\bibliography{Erfanianfar-MS-revised-2}

\begin{thebibliography}{}
\makeatletter
\relax
\def\mn@urlcharsother{\let\do\@makeother \do\$\do\&\do\#\do\^\do\_\do\%\do\~}
\def\mn@doi{\begingroup\mn@urlcharsother \@ifnextchar [ {\mn@doi@}
  {\mn@doi@[]}}
\def\mn@doi@[#1]#2{\def\@tempa{#1}\ifx\@tempa\@empty \href
  {http://dx.doi.org/#2} {doi:#2}\else \href {http://dx.doi.org/#2} {#1}\fi
  \endgroup}
\def\mn@eprint#1#2{\mn@eprint@#1:#2::\@nil}
\def\mn@eprint@arXiv#1{\href {http://arxiv.org/abs/#1} {{\tt arXiv:#1}}}
\def\mn@eprint@dblp#1{\href {http://dblp.uni-trier.de/rec/bibtex/#1.xml}
  {dblp:#1}}
\def\mn@eprint@#1:#2:#3:#4\@nil{\def\@tempa {#1}\def\@tempb {#2}\def\@tempc
  {#3}\ifx \@tempc \@empty \let \@tempc \@tempb \let \@tempb \@tempa \fi \ifx
  \@tempb \@empty \def\@tempb {arXiv}\fi \@ifundefined
  {mn@eprint@\@tempb}{\@tempb:\@tempc}{\expandafter \expandafter \csname
  mn@eprint@\@tempb\endcsname \expandafter{\@tempc}}}

\bibitem[\protect\citeauthoryear{{Abadi}, {Moore}  \& {Bower}}{{Abadi}
  et~al.}{1999}]{Abadi1999}
{Abadi} M.~G.,  {Moore} B.,   {Bower} R.~G.,  1999, \mn@doi [\mnras]
  {10.1046/j.1365-8711.1999.02715.x}, \href
  {http://adsabs.harvard.edu/abs/1999MNRAS.308..947A} {308, 947}

\bibitem[\protect\citeauthoryear{{Abramson}, {Kelson}, {Dressler}, {Poggianti},
  {Gladders}, {Oemler}  \& {Vulcani}}{{Abramson} et~al.}{2014}]{Abramson2014}
{Abramson} L.~E.,  {Kelson} D.~D.,  {Dressler} A.,  {Poggianti} B.,  {Gladders}
  M.~D.,  {Oemler} Jr. A.,   {Vulcani} B.,  2014, \mn@doi [\apjl]
  {10.1088/2041-8205/785/2/L36}, \href
  {http://adsabs.harvard.edu/abs/2014ApJ...785L..36A} {785, L36}

\bibitem[\protect\citeauthoryear{{Behroozi}, {Wechsler}  \&
  {Conroy}}{{Behroozi} et~al.}{2013}]{Behroozi2013}
{Behroozi} P.~S.,  {Wechsler} R.~H.,   {Conroy} C.,  2013, \mn@doi [\apj]
  {10.1088/0004-637X/770/1/57}, \href
  {http://adsabs.harvard.edu/abs/2013ApJ...770...57B} {770, 57}

\bibitem[\protect\citeauthoryear{{Birnboim} \& {Dekel}}{{Birnboim} \&
  {Dekel}}{2003}]{Birnboim2003}
{Birnboim} Y.,  {Dekel} A.,  2003, \mn@doi [\mnras]
  {10.1046/j.1365-8711.2003.06955.x}, \href
  {http://adsabs.harvard.edu/abs/2003MNRAS.345..349B} {345, 349}

\bibitem[\protect\citeauthoryear{{Blanton} et~al.,}{{Blanton}
  et~al.}{2005}]{Blanton2005}
{Blanton} M.~R.,  et~al., 2005, \mn@doi [\aj] {10.1086/429803}, \href
  {http://adsabs.harvard.edu/abs/2005AJ....129.2562B} {129, 2562}

\bibitem[\protect\citeauthoryear{{Bongiorno} et~al.,}{{Bongiorno}
  et~al.}{2012}]{Bongiorno2012}
{Bongiorno} A.,  et~al., 2012, \mn@doi [\mnras]
  {10.1111/j.1365-2966.2012.22089.x}, \href
  {http://adsabs.harvard.edu/abs/2012MNRAS.427.3103B} {427, 3103}

\bibitem[\protect\citeauthoryear{{Brinchmann}, {Charlot}, {White}, {Tremonti},
  {Kauffmann}, {Heckman}  \& {Brinkmann}}{{Brinchmann}
  et~al.}{2004}]{Brinchmann2004}
{Brinchmann} J.,  {Charlot} S.,  {White} S.~D.~M.,  {Tremonti} C.,  {Kauffmann}
  G.,  {Heckman} T.,   {Brinkmann} J.,  2004, \mn@doi [\mnras]
  {10.1111/j.1365-2966.2004.07881.x}, \href
  {http://adsabs.harvard.edu/abs/2004MNRAS.351.1151B} {351, 1151}

\bibitem[\protect\citeauthoryear{{Chabrier}}{{Chabrier}}{2003}]{Chabrier2003}
{Chabrier} G.,  2003, \mn@doi [\pasp] {10.1086/376392}, \href
  {http://adsabs.harvard.edu/abs/2003PASP..115..763C} {115, 763}

\bibitem[\protect\citeauthoryear{{Christlein} \& {Zabludoff}}{{Christlein} \&
  {Zabludoff}}{2005}]{Christlein2005}
{Christlein} D.,  {Zabludoff} A.~I.,  2005, \mn@doi [\apj] {10.1086/427427},
  \href {http://adsabs.harvard.edu/abs/2005ApJ...621..201C} {621, 201}

\bibitem[\protect\citeauthoryear{{Crocker} et~al.,}{{Crocker}
  et~al.}{2012}]{Crocker2012}
{Crocker} A.,  et~al., 2012, \mn@doi [\mnras]
  {10.1111/j.1365-2966.2011.20393.x}, \href
  {http://adsabs.harvard.edu/abs/2012MNRAS.421.1298C} {421, 1298}

\bibitem[\protect\citeauthoryear{{Croton} et~al.,}{{Croton}
  et~al.}{2006}]{Croton2006}
{Croton} D.~J.,  et~al., 2006, \mn@doi [\mnras]
  {10.1111/j.1365-2966.2005.09675.x}, \href
  {http://adsabs.harvard.edu/abs/2006MNRAS.365...11C} {365, 11}

\bibitem[\protect\citeauthoryear{{Cucciati} et~al.,}{{Cucciati}
  et~al.}{2010}]{Cucciati2010}
{Cucciati} O.,  et~al., 2010, \mn@doi [\aap] {10.1051/0004-6361/200912585},
  \href {http://adsabs.harvard.edu/abs/2010A\%26A...524A...2C} {524, A2}

\bibitem[\protect\citeauthoryear{{Daddi} et~al.,}{{Daddi}
  et~al.}{2007}]{Daddi2007}
{Daddi} E.,  et~al., 2007, \mn@doi [\apj] {10.1086/521818}, \href
  {http://adsabs.harvard.edu/abs/2007ApJ...670..156D} {670, 156}

\bibitem[\protect\citeauthoryear{{De Lucia}, {Springel}, {White}, {Croton}  \&
  {Kauffmann}}{{De Lucia} et~al.}{2006}]{DeLucia2006}
{De Lucia} G.,  {Springel} V.,  {White} S.~D.~M.,  {Croton} D.,   {Kauffmann}
  G.,  2006, \mn@doi [\mnras] {10.1111/j.1365-2966.2005.09879.x}, \href
  {http://adsabs.harvard.edu/abs/2006MNRAS.366..499D} {366, 499}

\bibitem[\protect\citeauthoryear{{De Lucia}, {Weinmann}, {Poggianti},
  {Arag{\'o}n-Salamanca}  \& {Zaritsky}}{{De Lucia} et~al.}{2012}]{Delucia2012}
{De Lucia} G.,  {Weinmann} S.,  {Poggianti} B.~M.,  {Arag{\'o}n-Salamanca} A.,
   {Zaritsky} D.,  2012, \mn@doi [\mnras] {10.1111/j.1365-2966.2012.20983.x},
  \href {http://adsabs.harvard.edu/abs/2012MNRAS.423.1277D} {423, 1277}

\bibitem[\protect\citeauthoryear{{Dekel} \& {Woo}}{{Dekel} \&
  {Woo}}{2003}]{Dekel2003}
{Dekel} A.,  {Woo} J.,  2003, \mn@doi [\mnras]
  {10.1046/j.1365-8711.2003.06923.x}, \href
  {http://adsabs.harvard.edu/abs/2003MNRAS.344.1131D} {344, 1131}

\bibitem[\protect\citeauthoryear{{Dekel} et~al.,}{{Dekel}
  et~al.}{2009}]{Dekel2009}
{Dekel} A.,  et~al., 2009, \mn@doi [\nat] {10.1038/nature07648}, \href
  {http://adsabs.harvard.edu/abs/2009Natur.457..451D} {457, 451}

\bibitem[\protect\citeauthoryear{{Diemand}, {Kuhlen}  \& {Madau}}{{Diemand}
  et~al.}{2007}]{Diemand2007}
{Diemand} J.,  {Kuhlen} M.,   {Madau} P.,  2007, \mn@doi [\apj]
  {10.1086/520573}, \href {http://adsabs.harvard.edu/abs/2007ApJ...667..859D}
  {667, 859}

\bibitem[\protect\citeauthoryear{{Elbaz} et~al.,}{{Elbaz}
  et~al.}{2007}]{Elbaz2007}
{Elbaz} D.,  et~al., 2007, \mn@doi [\aap] {10.1051/0004-6361:20077525}, \href
  {http://adsabs.harvard.edu/abs/2007A\%26A...468...33E} {468, 33}

\bibitem[\protect\citeauthoryear{{Elbaz} et~al.,}{{Elbaz}
  et~al.}{2011}]{Elbaz2011}
{Elbaz} D.,  et~al., 2011, \mn@doi [\aap] {10.1051/0004-6361/201117239}, \href
  {http://adsabs.harvard.edu/abs/2011A\%26A...533A.119E} {533, A119}

\bibitem[\protect\citeauthoryear{{Erfanianfar} et~al.,}{{Erfanianfar}
  et~al.}{2014}]{Erfanianfar2014}
{Erfanianfar} G.,  et~al., 2014, \mn@doi [\mnras] {10.1093/mnras/stu1883},
  \href {http://adsabs.harvard.edu/abs/2014MNRAS.445.2725E} {445, 2725}

\bibitem[\protect\citeauthoryear{{Farouki} \& {Shapiro}}{{Farouki} \&
  {Shapiro}}{1981}]{Farouki1981}
{Farouki} R.,  {Shapiro} S.~L.,  1981, \mn@doi [\apj] {10.1086/158563}, \href
  {http://adsabs.harvard.edu/abs/1981ApJ...243...32F} {243, 32}

\bibitem[\protect\citeauthoryear{{Finoguenov} et~al.,}{{Finoguenov}
  et~al.}{2015}]{Finoguenov2015}
{Finoguenov} A.,  et~al., 2015, \mn@doi [\aap] {10.1051/0004-6361/201323053},
  \href {http://adsabs.harvard.edu/abs/2015A%26A...576A.130F} {576, A130}

\bibitem[\protect\citeauthoryear{{Genzel} et~al.,}{{Genzel}
  et~al.}{2014}]{Genzel2014}
{Genzel} R.,  et~al., 2014, \mn@doi [\apj] {10.1088/0004-637X/785/1/75}, \href
  {http://adsabs.harvard.edu/abs/2014ApJ...785...75G} {785, 75}

\bibitem[\protect\citeauthoryear{{Graham} \& {Driver}}{{Graham} \&
  {Driver}}{2005}]{Graham2005}
{Graham} A.~W.,  {Driver} S.~P.,  2005, \mn@doi [\pasa] {10.1071/AS05001},
  \href {http://adsabs.harvard.edu/abs/2005PASA...22..118G} {22, 118}

\bibitem[\protect\citeauthoryear{{Griffith} et~al.,}{{Griffith}
  et~al.}{2012}]{Griffith2012}
{Griffith} R.~L.,  et~al., 2012, \mn@doi [\apjs] {10.1088/0067-0049/200/1/9},
  \href {http://adsabs.harvard.edu/abs/2012ApJS..200....9G} {200, 9}

\bibitem[\protect\citeauthoryear{{Gunn} \& {Gott}}{{Gunn} \&
  {Gott}}{1972}]{Gunn1972}
{Gunn} J.~E.,  {Gott} I. J.~R.,  1972, \mn@doi [\apj] {10.1086/151605}, \href
  {http://adsabs.harvard.edu/abs/1972ApJ...176....1G} {176, 1}

\bibitem[\protect\citeauthoryear{{Harrison} et~al.,}{{Harrison}
  et~al.}{2012}]{Harrison2012}
{Harrison} C.~M.,  et~al., 2012, \mn@doi [\apjl] {10.1088/2041-8205/760/1/L15},
  \href {http://adsabs.harvard.edu/abs/2012ApJ...760L..15H} {760, L15}

\bibitem[\protect\citeauthoryear{{Hopkins}, {Hernquist}, {Cox}, {Di Matteo},
  {Robertson}  \& {Springel}}{{Hopkins} et~al.}{2006}]{Hopkins2006}
{Hopkins} P.~F.,  {Hernquist} L.,  {Cox} T.~J.,  {Di Matteo} T.,  {Robertson}
  B.,   {Springel} V.,  2006, \mn@doi [\apjs] {10.1086/499298}, \href
  {http://adsabs.harvard.edu/abs/2006ApJS..163....1H} {163, 1}

\bibitem[\protect\citeauthoryear{{Huang}, {Haynes}, {Giovanelli}  \&
  {Brinchmann}}{{Huang} et~al.}{2012}]{Huang2012}
{Huang} S.,  {Haynes} M.~P.,  {Giovanelli} R.,   {Brinchmann} J.,  2012,
  \mn@doi [\apj] {10.1088/0004-637X/756/2/113}, \href
  {http://adsabs.harvard.edu/abs/2012ApJ...756..113H} {756, 113}

\bibitem[\protect\citeauthoryear{{Ilbert} et~al.,}{{Ilbert}
  et~al.}{2010}]{Ilbert2010}
{Ilbert} O.,  et~al., 2010, \mn@doi [\apj] {10.1088/0004-637X/709/2/644}, \href
  {http://adsabs.harvard.edu/abs/2010ApJ...709..644I} {709, 644}

\bibitem[\protect\citeauthoryear{{Iovino} et~al.,}{{Iovino}
  et~al.}{2010}]{Iovino2010}
{Iovino} A.,  et~al., 2010, \mn@doi [\aap] {10.1051/0004-6361/200912558}, \href
  {http://adsabs.harvard.edu/abs/2010A\%26A...509A..40I} {509, A40}

\bibitem[\protect\citeauthoryear{{Kauffmann}, {White}, {Heckman}, {M{\'e}nard},
  {Brinchmann}, {Charlot}, {Tremonti}  \& {Brinkmann}}{{Kauffmann}
  et~al.}{2004}]{Kauffmann2004}
{Kauffmann} G.,  {White} S.~D.~M.,  {Heckman} T.~M.,  {M{\'e}nard} B.,
  {Brinchmann} J.,  {Charlot} S.,  {Tremonti} C.,   {Brinkmann} J.,  2004,
  \mn@doi [\mnras] {10.1111/j.1365-2966.2004.08117.x}, \href
  {http://adsabs.harvard.edu/abs/2004MNRAS.353..713K} {353, 713}

\bibitem[\protect\citeauthoryear{{Kennicutt}}{{Kennicutt}}{1998}]{Kennicutt1998}
{Kennicutt} J. R.~C.,  1998, \mn@doi [\araa] {10.1146/annurev.astro.36.1.189},
  \href {http://adsabs.harvard.edu/abs/1998ARA\%26A..36..189K} {36, 189}

\bibitem[\protect\citeauthoryear{{Kere\v{s}}, {Katz}, {Weinberg}  \&
  {Dav{\'e}}}{{Kere\v{s}} et~al.}{2005}]{Keres2005}
{Kere\v{s}} D.,  {Katz} N.,  {Weinberg} D.~H.,   {Dav{\'e}} R.,  2005, \mn@doi
  [\mnras] {10.1111/j.1365-2966.2005.09451.x}, \href
  {http://adsabs.harvard.edu/abs/2005MNRAS.363....2K} {363, 2}

\bibitem[\protect\citeauthoryear{{Kitzbichler} \& {White}}{{Kitzbichler} \&
  {White}}{2007}]{Kitzbichler2007}
{Kitzbichler} M.~G.,  {White} S.~D.~M.,  2007, \mn@doi [\mnras]
  {10.1111/j.1365-2966.2007.11458.x}, \href
  {http://adsabs.harvard.edu/abs/2007MNRAS.376....2K} {376, 2}

\bibitem[\protect\citeauthoryear{{Kova\v{c}} et~al.,}{{Kova\v{c}}
  et~al.}{2010}]{Kovac2010}
{Kova\v{c}} K.,  et~al., 2010, \mn@doi [\apj] {10.1088/0004-637X/718/1/86},
  \href {http://adsabs.harvard.edu/abs/2010ApJ...718...86K} {718, 86}

\bibitem[\protect\citeauthoryear{{Lang} et~al.,}{{Lang}
  et~al.}{2014}]{Lang2014}
{Lang} P.,  et~al., 2014, \mn@doi [\apj] {10.1088/0004-637X/788/1/11}, \href
  {http://adsabs.harvard.edu/abs/2014ApJ...788...11L} {788, 11}

\bibitem[\protect\citeauthoryear{{Larson}, {Tinsley}  \& {Caldwell}}{{Larson}
  et~al.}{1980}]{Larson1980}
{Larson} R.~B.,  {Tinsley} B.~M.,   {Caldwell} C.~N.,  1980, \mn@doi [\apj]
  {10.1086/157917}, \href {http://adsabs.harvard.edu/abs/1980ApJ...237..692L}
  {237, 692}

\bibitem[\protect\citeauthoryear{{Le Floc'h} et~al.,}{{Le Floc'h}
  et~al.}{2005}]{LeFloch2005}
{Le Floc'h} E.,  et~al., 2005, \mn@doi [\apj] {10.1086/432789}, \href
  {http://adsabs.harvard.edu/abs/2005ApJ...632..169L} {632, 169}

\bibitem[\protect\citeauthoryear{{Lee} et~al.,}{{Lee} et~al.}{2015}]{Lee2015}
{Lee} N.,  et~al., 2015, \mn@doi [\apj] {10.1088/0004-637X/801/2/80}, \href
  {http://adsabs.harvard.edu/abs/2015ApJ...801...80L} {801, 80}

\bibitem[\protect\citeauthoryear{{Leitner}}{{Leitner}}{2012}]{Leitner2012}
{Leitner} S.~N.,  2012, \mn@doi [\apj] {10.1088/0004-637X/745/2/149}, \href
  {http://adsabs.harvard.edu/abs/2012ApJ...745..149L} {745, 149}

\bibitem[\protect\citeauthoryear{{Leja}, {van Dokkum}, {Franx}  \&
  {Whitaker}}{{Leja} et~al.}{2015}]{Leja2015}
{Leja} J.,  {van Dokkum} P.~G.,  {Franx} M.,   {Whitaker} K.~E.,  2015, \mn@doi
  [\apj] {10.1088/0004-637X/798/2/115}, \href
  {http://adsabs.harvard.edu/abs/2015ApJ...798..115L} {798, 115}

\bibitem[\protect\citeauthoryear{{Lilly}, {Le Fevre}, {Hammer}  \&
  {Crampton}}{{Lilly} et~al.}{1996}]{Lilly1996}
{Lilly} S.~J.,  {Le Fevre} O.,  {Hammer} F.,   {Crampton} D.,  1996, \mn@doi
  [\apjl] {10.1086/309975}, \href
  {http://adsabs.harvard.edu/abs/1996ApJ...460L...1L} {460, L1}

\bibitem[\protect\citeauthoryear{{Madau}, {Pozzetti}  \& {Dickinson}}{{Madau}
  et~al.}{1998}]{Madau1998}
{Madau} P.,  {Pozzetti} L.,   {Dickinson} M.,  1998, \mn@doi [\apj]
  {10.1086/305523}, \href {http://adsabs.harvard.edu/abs/1998ApJ...498..106M}
  {498, 106}

\bibitem[\protect\citeauthoryear{{Martig}, {Bournaud}, {Teyssier}  \&
  {Dekel}}{{Martig} et~al.}{2009}]{Martig2009}
{Martig} M.,  {Bournaud} F.,  {Teyssier} R.,   {Dekel} A.,  2009, \mn@doi
  [\apj] {10.1088/0004-637X/707/1/250}, \href
  {http://adsabs.harvard.edu/abs/2009ApJ...707..250M} {707, 250}

\bibitem[\protect\citeauthoryear{{Moore}, {Governato}, {Quinn}, {Stadel}  \&
  {Lake}}{{Moore} et~al.}{1998}]{Moore1998}
{Moore} B.,  {Governato} F.,  {Quinn} T.,  {Stadel} J.,   {Lake} G.,  1998,
  \mn@doi [\apjl] {10.1086/311333}, \href
  {http://adsabs.harvard.edu/abs/1998ApJ...499L...5M} {499, L5}

\bibitem[\protect\citeauthoryear{{Mullaney} et~al.,}{{Mullaney}
  et~al.}{2012}]{Mullaney2012}
{Mullaney} J.~R.,  et~al., 2012, \mn@doi [\apjl] {10.1088/2041-8205/753/2/L30},
  \href {http://adsabs.harvard.edu/abs/2012ApJ...753L..30M} {753, L30}

\bibitem[\protect\citeauthoryear{{Noeske} et~al.,}{{Noeske}
  et~al.}{2007}]{Noeske2007a}
{Noeske} K.~G.,  et~al., 2007, \mn@doi [\apjl] {10.1086/517926}, \href
  {http://adsabs.harvard.edu/abs/2007ApJ...660L..43N} {660, L43}

\bibitem[\protect\citeauthoryear{{Park} \& {Hwang}}{{Park} \&
  {Hwang}}{2009}]{Park2009}
{Park} C.,  {Hwang} H.~S.,  2009, \mn@doi [\apj]
  {10.1088/0004-637X/699/2/1595}, \href
  {http://adsabs.harvard.edu/abs/2009ApJ...699.1595P} {699, 1595}

\bibitem[\protect\citeauthoryear{{Peng} et~al.,}{{Peng}
  et~al.}{2010}]{Peng2010}
{Peng} Y.,  et~al., 2010, \mn@doi [\apj] {10.1088/0004-637X/721/1/193}, \href
  {http://adsabs.harvard.edu/abs/2010ApJ...721..193P} {721, 193}

\bibitem[\protect\citeauthoryear{{Peng}, {Lilly}, {Renzini}  \&
  {Carollo}}{{Peng} et~al.}{2012}]{Peng2012}
{Peng} Y.,  {Lilly} S.~J.,  {Renzini} A.,   {Carollo} M.,  2012, \mn@doi [\apj]
  {10.1088/0004-637X/757/1/4}, \href
  {http://adsabs.harvard.edu/abs/2012ApJ...757....4P} {757, 4}

\bibitem[\protect\citeauthoryear{{Poggianti} et~al.,}{{Poggianti}
  et~al.}{2008}]{Poggianti2008}
{Poggianti} B.~M.,  et~al., 2008, \mn@doi [\apj] {10.1086/589936}, \href
  {http://adsabs.harvard.edu/abs/2008ApJ...684..888P} {684, 888}

\bibitem[\protect\citeauthoryear{{Poglitsch} et~al.,}{{Poglitsch}
  et~al.}{2010}]{Poglitsch2010}
{Poglitsch} A.,  et~al., 2010, \mn@doi [\aap] {10.1051/0004-6361/201014535},
  \href {http://adsabs.harvard.edu/abs/2010A\%26A...518L...2P} {518, L2}

\bibitem[\protect\citeauthoryear{{Popesso} et~al.,}{{Popesso}
  et~al.}{2015a}]{Popesso2015a}
{Popesso} P.,  et~al., 2015a, \mn@doi [\aap] {10.1051/0004-6361/201424711},
  \href {http://adsabs.harvard.edu/abs/2015A%26A...574A.105P} {574, A105}

\bibitem[\protect\citeauthoryear{{Popesso} et~al.,}{{Popesso}
  et~al.}{2015b}]{Popesso2015b}
{Popesso} P.,  et~al., 2015b, \mn@doi [\aap] {10.1051/0004-6361/201424715},
  \href {http://adsabs.harvard.edu/abs/2015A%26A...579A.132P} {579, A132}

\bibitem[\protect\citeauthoryear{{Quilis}, {Bower}  \& {Balogh}}{{Quilis}
  et~al.}{2001}]{Quilis2001}
{Quilis} V.,  {Bower} R.~G.,   {Balogh} M.~L.,  2001, \mn@doi [\mnras]
  {10.1046/j.1365-8711.2001.04927.x}, \href
  {http://adsabs.harvard.edu/abs/2001MNRAS.328.1091Q} {328, 1091}

\bibitem[\protect\citeauthoryear{{Rodighiero} et~al.,}{{Rodighiero}
  et~al.}{2010}]{Rodighiero2010}
{Rodighiero} G.,  et~al., 2010, \mn@doi [\aap] {10.1051/0004-6361/201014624},
  \href {http://adsabs.harvard.edu/abs/2010A\%26A...518L..25R} {518, L25}

\bibitem[\protect\citeauthoryear{{Rosario} et~al.,}{{Rosario}
  et~al.}{2012}]{Rosario2012}
{Rosario} D.~J.,  et~al., 2012, \mn@doi [\aap] {10.1051/0004-6361/201219258},
  \href {http://adsabs.harvard.edu/abs/2012A\%26A...545A..45R} {545, A45}

\bibitem[\protect\citeauthoryear{{Rovilos} et~al.,}{{Rovilos}
  et~al.}{2012}]{Rovilos2012}
{Rovilos} E.,  et~al., 2012, \mn@doi [\aap] {10.1051/0004-6361/201218952},
  \href {http://adsabs.harvard.edu/abs/2012A%26A...546A..58R} {546, A58}

\bibitem[\protect\citeauthoryear{{Saintonge} et~al.,}{{Saintonge}
  et~al.}{2012}]{Saintonge2012}
{Saintonge} A.,  et~al., 2012, \mn@doi [\apj] {10.1088/0004-637X/758/2/73},
  \href {http://adsabs.harvard.edu/abs/2012ApJ...758...73S} {758, 73}

\bibitem[\protect\citeauthoryear{{Salim} et~al.,}{{Salim}
  et~al.}{2007}]{Salim2007}
{Salim} S.,  et~al., 2007, \mn@doi [\apjs] {10.1086/519218}, \href
  {http://adsabs.harvard.edu/abs/2007ApJS..173..267S} {173, 267}

\bibitem[\protect\citeauthoryear{{Scodeggio} et~al.,}{{Scodeggio}
  et~al.}{2009}]{Scodeggio2009}
{Scodeggio} M.,  et~al., 2009, \mn@doi [\aap] {10.1051/0004-6361/200810511},
  \href {http://adsabs.harvard.edu/abs/2009A\%26A...501...21S} {501, 21}

\bibitem[\protect\citeauthoryear{{Sersic}, {Garcia Lambas}  \&
  {Mosconi}}{{Sersic} et~al.}{1986}]{Sersic1986}
{Sersic} J.~L.,  {Garcia Lambas} D.,   {Mosconi} M.~B.,  1986, \rmxaa, \href
  {http://adsabs.harvard.edu/abs/1986RMxAA..12..132S} {12, 132}

\bibitem[\protect\citeauthoryear{{Simha}, {Weinberg}, {Dav{\'e}}, {Gnedin},
  {Katz}  \& {Kere{\v s}}}{{Simha} et~al.}{2009}]{Simha2009}
{Simha} V.,  {Weinberg} D.~H.,  {Dav{\'e}} R.,  {Gnedin} O.~Y.,  {Katz} N.,
  {Kere{\v s}} D.,  2009, \mn@doi [\mnras] {10.1111/j.1365-2966.2009.15341.x},
  \href {http://adsabs.harvard.edu/abs/2009MNRAS.399..650S} {399, 650}

\bibitem[\protect\citeauthoryear{{Somerville}, {Hopkins}, {Cox}, {Robertson}
  \& {Hernquist}}{{Somerville} et~al.}{2008}]{Somerville2008}
{Somerville} R.~S.,  {Hopkins} P.~F.,  {Cox} T.~J.,  {Robertson} B.~E.,
  {Hernquist} L.,  2008, \mn@doi [\mnras] {10.1111/j.1365-2966.2008.13805.x},
  \href {http://adsabs.harvard.edu/abs/2008MNRAS.391..481S} {391, 481}

\bibitem[\protect\citeauthoryear{{Speagle}, {Steinhardt}, {Capak}  \&
  {Silverman}}{{Speagle} et~al.}{2014}]{Speagle2014}
{Speagle} J.~S.,  {Steinhardt} C.~L.,  {Capak} P.~L.,   {Silverman} J.~D.,
  2014, \mn@doi [\apjs] {10.1088/0067-0049/214/2/15}, \href
  {http://adsabs.harvard.edu/abs/2014ApJS..214...15S} {214, 15}

\bibitem[\protect\citeauthoryear{{Tasca} et~al.,}{{Tasca}
  et~al.}{2009}]{Tasca2009}
{Tasca} L.~A.~M.,  et~al., 2009, \mn@doi [\aap] {10.1051/0004-6361/200912213},
  \href {http://adsabs.harvard.edu/abs/2009A\%26A...503..379T} {503, 379}

\bibitem[\protect\citeauthoryear{{Weinmann}, {van den Bosch}, {Yang}  \&
  {Mo}}{{Weinmann} et~al.}{2006}]{Weinmann2006}
{Weinmann} S.~M.,  {van den Bosch} F.~C.,  {Yang} X.,   {Mo} H.~J.,  2006,
  \mn@doi [\mnras] {10.1111/j.1365-2966.2005.09865.x}, \href
  {http://adsabs.harvard.edu/abs/2006MNRAS.366....2W} {366, 2}

\bibitem[\protect\citeauthoryear{{Weinmann}, {Kauffmann}, {van den Bosch},
  {Pasquali}, {McIntosh}, {Mo}, {Yang}  \& {Guo}}{{Weinmann}
  et~al.}{2009}]{Weinmann2009}
{Weinmann} S.~M.,  {Kauffmann} G.,  {van den Bosch} F.~C.,  {Pasquali} A.,
  {McIntosh} D.~H.,  {Mo} H.,  {Yang} X.,   {Guo} Y.,  2009, \mn@doi [\mnras]
  {10.1111/j.1365-2966.2009.14412.x}, \href
  {http://adsabs.harvard.edu/abs/2009MNRAS.394.1213W} {394, 1213}

\bibitem[\protect\citeauthoryear{{Wetzel}, {Tinker}, {Conroy}  \& {van den
  Bosch}}{{Wetzel} et~al.}{2013}]{Wetzel2013}
{Wetzel} A.~R.,  {Tinker} J.~L.,  {Conroy} C.,   {van den Bosch} F.~C.,  2013,
  \mn@doi [\mnras] {10.1093/mnras/stt469}, \href
  {http://adsabs.harvard.edu/abs/2013MNRAS.432..336W} {432, 336}

\bibitem[\protect\citeauthoryear{{Wheeler}, {Phillips}, {Cooper},
  {Boylan-Kolchin}  \& {Bullock}}{{Wheeler} et~al.}{2014}]{Wheeler2014}
{Wheeler} C.,  {Phillips} J.~I.,  {Cooper} M.~C.,  {Boylan-Kolchin} M.,
  {Bullock} J.~S.,  2014, \mn@doi [\mnras] {10.1093/mnras/stu965}, \href
  {http://adsabs.harvard.edu/abs/2014MNRAS.442.1396W} {442, 1396}

\bibitem[\protect\citeauthoryear{{Whitaker}, {van Dokkum}, {Brammer}  \&
  {Franx}}{{Whitaker} et~al.}{2012}]{Whitaker2012}
{Whitaker} K.~E.,  {van Dokkum} P.~G.,  {Brammer} G.,   {Franx} M.,  2012,
  \mn@doi [\apjl] {10.1088/2041-8205/754/2/L29}, \href
  {http://adsabs.harvard.edu/abs/2012ApJ...754L..29W} {754, L29}

\bibitem[\protect\citeauthoryear{{Whitaker} et~al.,}{{Whitaker}
  et~al.}{2014}]{Whitaker2014}
{Whitaker} K.~E.,  et~al., 2014, \mn@doi [\apj] {10.1088/0004-637X/795/2/104},
  \href {http://adsabs.harvard.edu/abs/2014ApJ...795..104W} {795, 104}

\bibitem[\protect\citeauthoryear{{Whitaker} et~al.,}{{Whitaker}
  et~al.}{2015}]{Whitaker2015}
{Whitaker} K.~E.,  et~al., 2015, ArXiv e-prints/508.04771, \href
  {http://adsabs.harvard.edu/abs/2015arXiv150804771W} {}

\bibitem[\protect\citeauthoryear{{Wuyts} et~al.,}{{Wuyts}
  et~al.}{2011}]{Wuyts2011}
{Wuyts} S.,  et~al., 2011, \mn@doi [\apj] {10.1088/0004-637X/742/2/96}, \href
  {http://adsabs.harvard.edu/abs/2011ApJ...742...96W} {742, 96}

\bibitem[\protect\citeauthoryear{{Ziparo} et~al.,}{{Ziparo}
  et~al.}{2013}]{Ziparo2013}
{Ziparo} F.,  et~al., 2013, \mn@doi [\mnras] {10.1093/mnras/stt1222}, \href
  {http://adsabs.harvard.edu/abs/2013MNRAS.434.3089Z} {434, 3089}

\bibitem[\protect\citeauthoryear{{Ziparo} et~al.,}{{Ziparo}
  et~al.}{2014}]{Ziparo2014}
{Ziparo} F.,  et~al., 2014, \mn@doi [\mnras] {10.1093/mnras/stt1901}, \href
  {http://adsabs.harvard.edu/abs/2014MNRAS.437..458Z} {437, 458}

\bibitem[\protect\citeauthoryear{{van den Bosch}, {Aquino}, {Yang}, {Mo},
  {Pasquali}, {McIntosh}, {Weinmann}  \& {Kang}}{{van den Bosch}
  et~al.}{2008}]{vandenBosch2008}
{van den Bosch} F.~C.,  {Aquino} D.,  {Yang} X.,  {Mo} H.~J.,  {Pasquali} A.,
  {McIntosh} D.~H.,  {Weinmann} S.~M.,   {Kang} X.,  2008, \mn@doi [\mnras]
  {10.1111/j.1365-2966.2008.13230.x}, \href
  {http://adsabs.harvard.edu/abs/2008MNRAS.387...79V} {387, 79}

\makeatother
\end{thebibliography}
\appendix
 \section{Error analysis}\label{App:AppendixA}
 
\begin{figure}
\centering
\includegraphics[width=8cm]{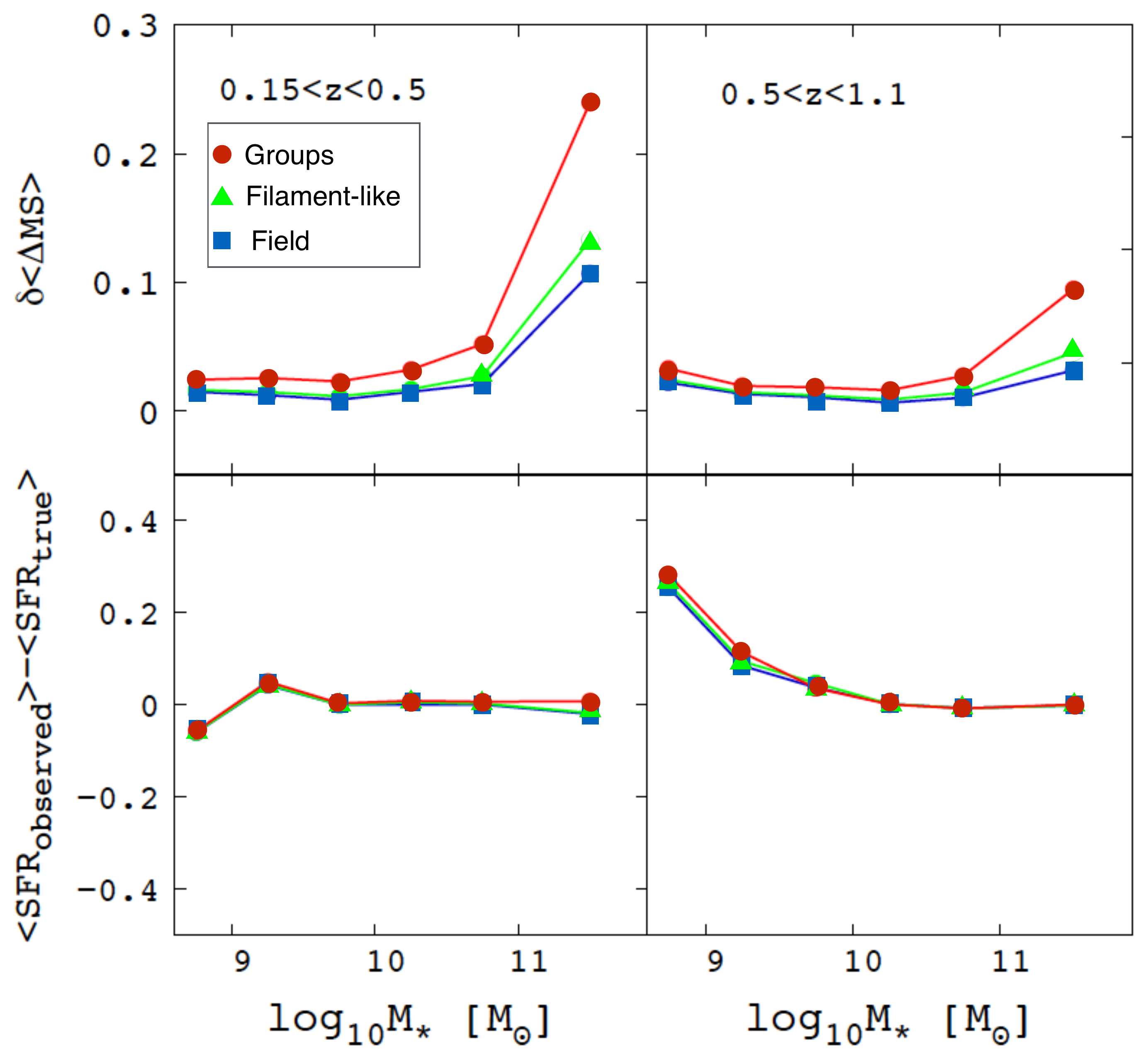}
\caption[Dispersion around the MS location as a function of the galaxy stellar mass]{{\it upper panels:} Dispersion around the MS$_{complete}$-MS$_{incomplete}$ for simulated group galaxies (in red), filament-like galaxies (in green) and field
galaxies(in blue) as a function of stellar mass. {\it lower panels:} Deviation of the simulated MS galaxies  in different environment form the complete original Mock catalog as a function of stellar mass.}
\label{deltadelta}
\end{figure}

 In order to estimate the errors in our analysis, we use the mock catalog of \cite{Kitzbichler2007}.  Using the available photometric bands (RJ ),
  we simulate the spectroscopic selection function of the surveys used in this work by extracting randomly in each magnitude bin a percentage
   of galaxies consistent with the percentage of systems with spectroscopic redshift in the same magnitude bin observed in each of our surveys.
    We do this separately for each survey, since each field shows a different spectroscopic selection function (see Figure 5 in E14). We use this analysis to check for biases due to the spectroscopic selection function and to estimate the error due to the low number statistics. 
 We create randomly 5000 ``incomplete'' catalogs in this way in the low and high redshift bins. The dispersion around the difference between MS obtained from original complete catalog (MS$_{complete}$) and the one from our simulation (MS$_{incomplete}$) for each environment provides the error on the $\Delta$MS for that environment. 
  The upper panels of Figure \ref{deltadelta} shows the error in different stellar mass bins for different environments in our two redshift bins. In order to see if our analysis are biased, we also look at the deviation of  the MS$_{incomplete}$ from MS$_{complete}$. The lower panels of Figure \ref{deltadelta} show that in high stellar masses (above 10$^{9.5})$, there is no bias due to high spectroscopic completeness. However, there is a bias at lower stellar masses which are not affecting our results. As the errors obtained in this way is lower than the standard errors, we keep the standard errors as the errors on $\Delta$MS.

\section{Wavelength Dependence of Morphological Measurements} \label{App:AppendixB}
\begin{figure}
 \centering
\includegraphics[width=0.4\textwidth]{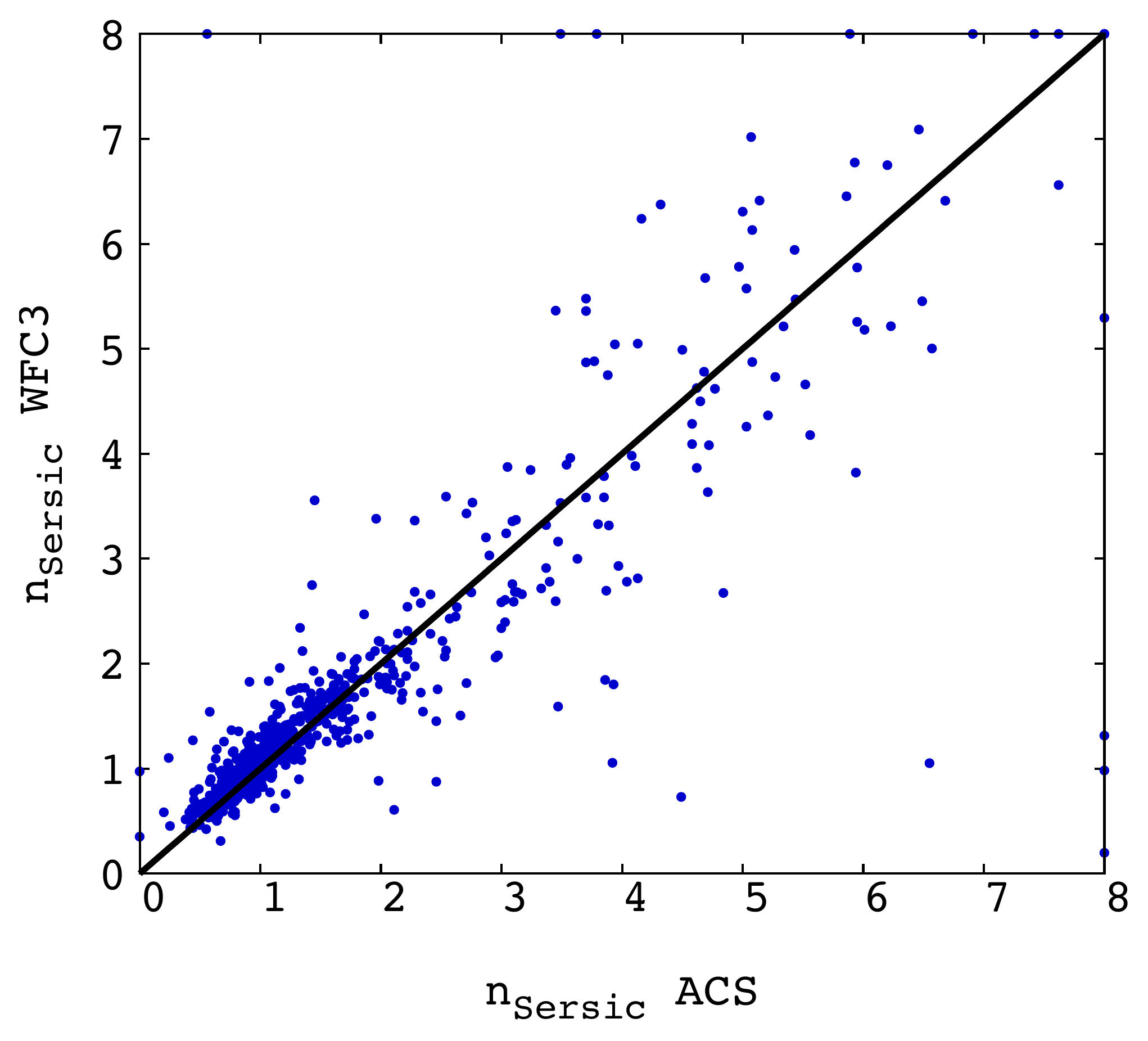}
\includegraphics[width=0.4\textwidth]{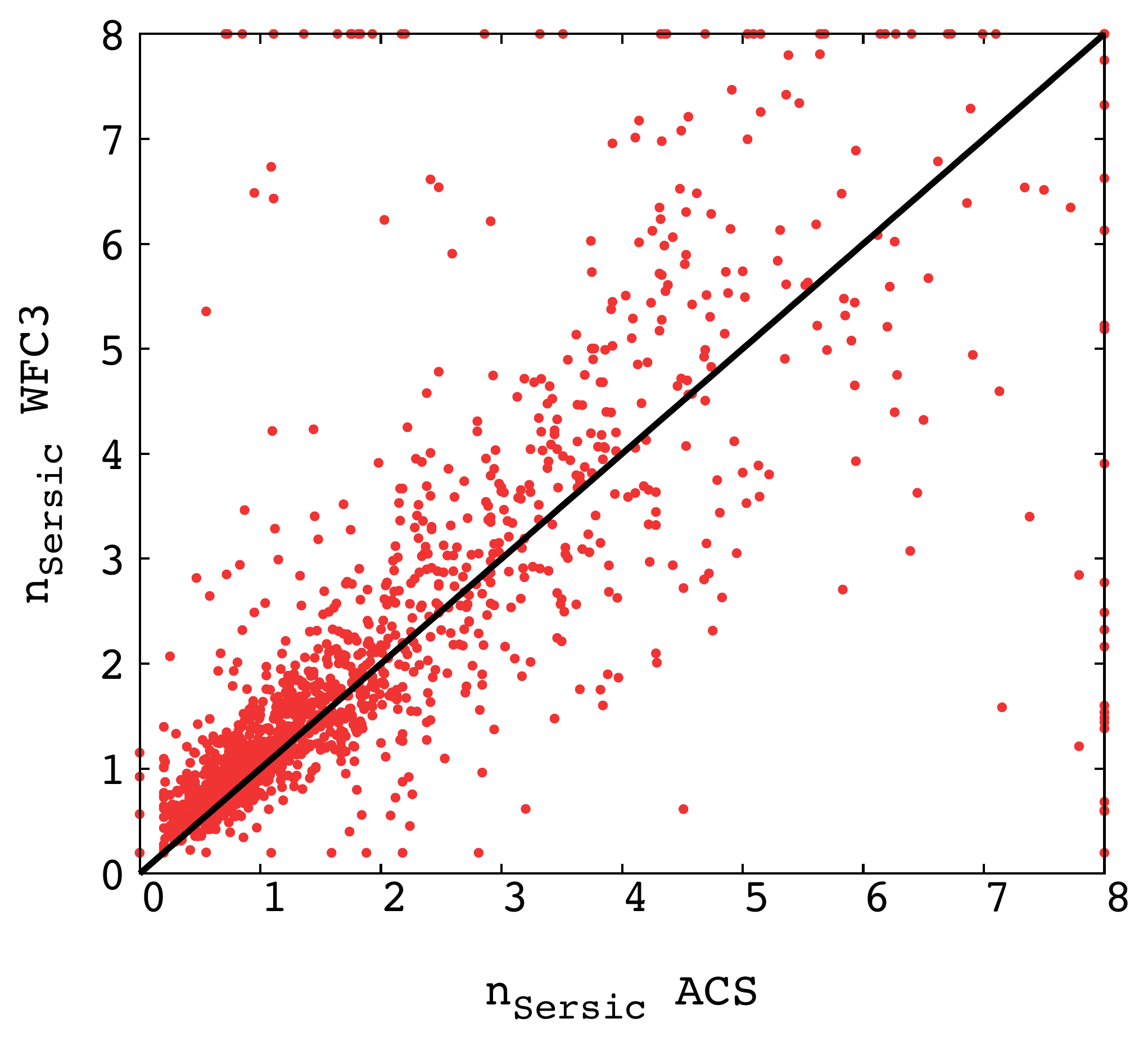}
\caption[SFR-stellar mass relation for field, filament-like and group galaxies]{The S\'ersic index measured on WFC3 F125W band vs. ACS F850LP and ACS F814W bands in GOODS-S and COSMOS, respectively. The upper panel corresponds to the lows sample and the lower panel shows the highz one. }
\label{compare}
\end{figure}

Figure \ref{compare} compares measurements of S\'ersic index carried out on the WFC3 F215W mosaic and on the ACS F850LP and ACS F814W mosaics in GOODS-S and COSMOS, respectively. Median offsets are limited to the few percent level for both lowz and highz sample (2\% and 8\%, respectively) and the dispersion around the one to one relation is about 18\% for lowz sample and 30\% for the highz one.

\end{document}